\newif\iffull\fulltrue

\documentclass[twoside,leqno,twocolumn]{article}

\usepackage[letterpaper]{geometry}

\usepackage{ltexpprt}
\usepackage{hyperref}

\usepackage[utf8]{inputenc}

\usepackage{pgfplots}

\usepackage{amsmath, amsfonts, amssymb}
\usepackage[square,numbers]{natbib}
\usepackage{cleveref}
\usepackage{xy}
\usepackage{tikz}
\usepackage{mathrsfs}
\usepackage{setspace,dsfont,graphicx,makecell,appendix,multirow,float}
\pgfplotsset{compat=1.17}

\usepackage{bbm}
\usepackage{graphicx}

\usepackage{times}
\usepackage{tabularx}
\usepackage{dblfloatfix}
\usepackage{xspace}
\usepackage{algorithm}
\usepackage[noend]{algpseudocode}
\usepackage{subcaption,graphicx,xurl,multicol,mathtools}
\usepackage{paralist}
\usepackage{color}
\usepackage{arydshln}
\usepackage[most]{tcolorbox}
\usepackage{stmaryrd}
\usepackage{booktabs}
\usepackage{siunitx}
\usepackage{bm}
\usepackage{enumitem}
\setlist{nosep}
\graphicspath{ {./figures/} }

\newtheorem{definition}{Definition}

\newcommand{\BigO}[1]{\ensuremath{O(#1)}}

\newcommand{\whp}[1]{w.h.p.}
\newcommand{\algname}[1]{\textnormal{\textsc{#1}}}

\algblock{ParFor}{EndParFor}
\algnewcommand\algorithmicparfor{\textbf{parfor}}
\algnewcommand\algorithmicpardo{\textbf{do}}
\algnewcommand\algorithmicendparfor{}%
\algrenewtext{ParFor}[1]{\algorithmicparfor\ #1\ \algorithmicpardo}
\algrenewtext{EndParFor}{\algorithmicendparfor}
\algtext*{EndParFor}

\algdef{SE}[DOWHILE]{Do}{doWhile}{\algorithmicdo}[1]{\algorithmicwhile\ #1}%

\newcommand{\myparagraph}[1]{\vspace{3pt}\noindent {\bf #1.}}
\usepackage{comment}

\newcommand{\kdt}{$k$d-tree\xspace}
\newcommand{\defn}[1]{\emph{\textbf{#1}}}

\newcommand{\savespace}[1]{}

\newcommand*\samethanks[1][\value{footnote}]{\footnotemark[#1]}

\usepackage[compact]{titlesec}
\titlespacing*{\section}{0pt}{*1}{*1}
\titlespacing*{\subsection}{0pt}{*1}{*1}
\titlespacing*{\subsubsection}{0pt}{*1}{*1}

\usepackage{times,microtype}

\begin{document}

\newcommand\relatedversion{}

\title{\Large Faster Parallel Exact Density Peaks Clustering\relatedversion{}}
\author{Yihao Huang\thanks{Phillips Academy} \ \thanks{The first two authors contributed equally to this work.}
\and Shangdi Yu\thanks{MIT CSAIL} \ \samethanks[2]
\and Julian Shun\samethanks[3]}

\date{}

\maketitle

\fancyfoot[R]{\scriptsize{Copyright \textcopyright\ 2023 by SIAM\\
Unauthorized reproduction of this article is prohibited}}

\begin{abstract}
\small\baselineskip=9pt 
Clustering multidimensional points is a fundamental data mining task, with applications in many fields, such as astronomy, neuroscience, bioinformatics, and computer vision. The goal of clustering algorithms is to group similar objects together. Density-based clustering is a clustering approach that defines clusters as dense regions of points. It has the advantage of being able to detect clusters of arbitrary shapes, rendering it useful in many applications. 

In this paper, we propose fast parallel algorithms for Density Peaks Clustering (DPC), a popular version of density-based clustering.
Existing exact DPC algorithms suffer from 
low parallelism both in theory and in practice, which limits their application to large-scale data sets. 
Our most performant algorithm, which is based on priority search \kdt{s}, achieves $\BigO{\log n\log\log n}$ span (parallel time complexity) for a data set of $n$ points.
Our algorithm is also work-efficient, achieving a work complexity matching the best existing sequential exact DPC algorithm. 
In addition, we present another DPC algorithm based on a Fenwick tree that makes fewer assumptions for its average-case complexity to hold.

We provide optimized implementations of our algorithms and evaluate their performance via extensive experiments. On a 30-core machine with two-way hyperthreading, we find that our best algorithm achieves a 10.8--13169x speedup over the previous best parallel exact DPC algorithm. 
Compared to the state-of-the-art parallel approximate DPC algorithm, our best algorithm achieves a 1.5--4206x speedup, while being exact. 

\end{abstract}

\section{Introduction}\label{sec:intro}
Clustering is the task of grouping similar objects into clusters and it is a fundamental task in data analysis and unsupervised machine learning. Clustering algorithms can be used to identify different types of tissues in medical imaging~\cite{Yang02}, analyze social networks, and identify weather regimes in climatology~\cite{Coe21}. They are also widely used as a data processing subroutine in other machine learning tasks~\cite{Coleman79,Wu22,Lin19,Marco13}. One popular type of clustering is density-based clustering, which defines clusters as dense regions of points in the coordinate space. Density-based clustering algorithms have received a lot of attention~\cite{Ester96,Agrawal98,Ankerst99,Januzaj04,Rodriguez14,Wang97,Hinneburg98,Hanmanthu18,Sheikholeslami00}, because they can discover clusters of arbitrary shapes, while many other popular algorithms, such as $k$-means, can only recover separable clusters with spherical shapes.

Density peak clustering (DPC)~\cite{Rodriguez14} is a popular version of density-based clustering. In this paper, we present fast parallel exact algorithms for DPC that outperform existing state-of-the-art implementations.
Many density-based clustering algorithms, such as DBSCAN~\cite{Ester96}, are sensitive towards the choice of a density-noise cutoff hyper-parameter (points with density lower than the cutoff are deemed as irrelevant noise)~\cite{Ester96}. DPC, in comparison, has been shown to perform well consistently over different hyper-parameter choices~\cite{Rodriguez14}. It is also very easy to set the hyper-parameters of DPC because DPC can generate a decision graph~\cite{Rodriguez14} that visually aids the determination of the hyper-parameters. Due to its advantages, DPC has been applied in many situations, such as the analysis of pathogenesis of COVID-19~\cite{Ziegler20}, cancer studies ~\cite{Guo18}, neuroscience studies~\cite{Peters07}, market analysis~\cite{Wang16-3}, computer vision tasks~\cite{Li18}, and natural language processing~\cite{Wang16}. 
DPC has three main steps:
\begin{list}{\arabic{enumi}.}{%
    \usecounter{enumi}
    \setlength{\leftmargin}{0.5em}
    \setlength{\itemindent}{0em}
    \setlength{\labelwidth}{\itemindent}
    \setlength{\labelsep}{0.5em}
    \setlength{\listparindent}{1em}
    \setlength{\itemsep}{0em}
    \setlength{\parsep}{0em}
    \setlength{\topsep}{0em}
    \setlength{\partopsep}{0em}
}
     \item Compute the density of each point $x$, which is the number of points in a ball centered at $x$ with a user-input parameter radius, $d_\text{cut}$. \label{item:density}
    \item For each point $x$, connect $x$ to its dependent point, which is the closest neighbor of $x$ that has a higher density than $x$. The resulting graph is a tree. \label{item:dep-point}
    \item Remove all connections with a distance higher than a certain threshold value. Each resulting connected component is a separate cluster. This final step is equivalent to performing single-linkage clustering~\cite{Rohlf82} on the tree. \label{item:single-linkage}
\end{list}

For a data set of $n$ points, a naive implementation of DPC that computes all pairwise point distances takes $\Theta(n^2)$ work to compute the density of all points and to connect each point to its dependent point~\cite{Rodriguez14}. This is expensive when the data set is large. 
Hence, multiple works have attempted to optimize the computational cost of DPC~\cite{Bai17, Xu21, Gong16, Zhang16, Rasool20, Amagata21}.
\citet{Rasool20} propose a sequential algorithm with average-case work complexity of $\BigO{n\log n}$. 
\citet{Amagata21} proposed a parallel algorithm that uses a \kdt to compute density values and find dependent points; it is currently the state-of-the-art parallel DPC algorithm for exact DPC clustering. Their algorithm is able to achieve the same average-case work complexity as \citet{Rasool20}'s algorithm, and their worst-case span\footnote{The \emph{span} is the length of the longest chain of sequential dependencies in the algorithm.} complexity is $\BigO{n\log n}$.\footnote{Amagata and Hara~\cite{Amagata21}'s implementation has a $\BigO{n^2}$ span complexity, but it can be trivially reduced to a $\BigO{n\log n}$ span by parallelizing the \kdt nearest neighbor search.} We will describe more about related work in \Cref{sec:related}.

As the sizes of modern data sets increase, it is important for clustering algorithms to have high parallelism, and ideally have polylogarithmic span. In this work, we develop parallel DPC algorithms to improve the span complexity of existing DPC algorithms. 
Our algorithms are able to achieve $\BigO{\log n\log\log n}$ worst case span complexity.  We present new parallel algorithms for Step~\ref{item:dep-point}---the parallelism bottleneck for \citet{Amagata21}'s algorithm. Our algorithms significantly reduce Step~\ref{item:dep-point}'s span complexity. We also optimize existing parallel algorithms for Steps~\ref{item:density} and \ref{item:single-linkage}.

Our first new algorithm for solving Step~\ref{item:dep-point}'s dependent point finding task utilizes a \defn{priority search \kdt}. A priority search \kdt is an optimization of a max \kdt \cite{Gro07,Duvenhage09}. The priority search \kdt can be constructed from a data set of $n$ points similar to a regular \kdt~\cite{wang2022esa} in $\BigO{n\log n}$ work and $\BigO{\log n\log\log{n}}$ span. A priority search \kdt can be used to directly find the dependent point of a point $x$. It is designed to process queries for the nearest neighbor of $x$ with a higher density value than $x$ itself. Hence, to compute the dependent point for every point in the data set, we only need to perform priority search \kdt queries for every point in the data set in parallel. The parallelism across different priority search \kdt queries allows our algorithm to avoid sequentiality in \citet{Amagata21}'s algorithm, achieving $\BigO{\log n}$ span complexity for finding dependent points while maintaining $\BigO{n\log n}$ average-case work. Since the priority search \kdt construction has a span complexity of $\BigO{\log n \log\log n}$, the overall span complexity of the algorithm is $\BigO{\log n \log\log n}$.

We also present a parallel Fenwick tree-based algorithm for finding dependent points. This algorithm stores points in multiple \kdt{s} nested inside a Fenwick tree. The Fenwick tree partitions points along increasing density values such that each \kdt{} stores points within a particular range of density values. To query the dependent point of a point $x$, we consider the range of density values higher than $x$'s density. This range is partitioned by the Fenwick tree into $\BigO{\log n}$ sub-ranges that each correspond to a \kdt. We perform queries on these $\BigO{\log n}$ \kdt{s} and aggregate the results. The algorithm is highly parallel since each dependent point query can be performed independently. The algorithm takes $\BigO{n^2}$ work in the worst case, but its average-case work is $\BigO{n\log^2 n}$. The span is again bounded by $\BigO{\log n\log\log n}$. Although this algorithm has a higher average-case work bound than our priority-search tree algorithm, its average-case complexity result requires fewer assumptions and it can sometimes be faster in practice.

In addition to our two dependent point finding algorithms, we also introduce an optimization technique for density computation: while counting the number of points within the neighborhood of a particular query point, we prune the searches through \kdt subtrees completely contained within that neighborhood by storing the number of points each subtree contains inside the \kdt and directly adding that number to the total number of points. 
Finally, we solve the single-linkage clustering step of the algorithm (Step ~\ref{item:single-linkage}) by using a parallel union-find data structure \cite{Jayanti21}, which has $\BigO{n\alpha(n,n)}$ expected work and $\BigO{\log n}$ span with high probability, where $\alpha$ represents the inverse Ackermann's function.  

We implement our algorithms and evaluate them on both synthetic and real-world data sets. We compare our runtime results to 
state-of-the-art exact and approximate DPC algorithms~\cite{Amagata21, Rasool20} on a 30-core machine with two-way hyper-threading. Our optimized density computation algorithm outperforms state-of-the-art~\cite{Amagata21} parallel exact density computation by 1.4--18586.3x. For dependent point finding, our parallel Fenwick tree based algorithm achieves 12.9--1551.7x speedup over state-of-the-art~\cite{Amagata21} algorithm and our parallel priority search \kdt based approach attains 8.3--4666.3x speedup. Considering the overall runtime, our best algorithm achieves a 10.8--13169x speedup over the previous best parallel exact DPC algorithm. Since our algorithms are exact, they give the same clustering quality as the original DPC algorithm. Compared to the state-of-the-art parallel approximate DPC algorithm, our best algorithm achieves a 1.5--4206x speedup, while being exact.

Our contributions are threefold:
\begin{list}{\arabic{enumi}.}{%
    \usecounter{enumi}
    \setlength{\leftmargin}{0.5em}
    \setlength{\itemindent}{0em}
    \setlength{\labelwidth}{\itemindent}
    \setlength{\labelsep}{0.5em}
    \setlength{\listparindent}{1em}
    \setlength{\itemsep}{0em}
    \setlength{\parsep}{0em}
    \setlength{\topsep}{0em}
    \setlength{\partopsep}{0em}
}
    \item We introduce two novel algorithms for solving the dependent point finding task in DPC, and introduce techniques for speeding up the density computation and single-linkage clustering tasks in DPC. 
    \item We provide theoretical bounds of our algorithms as well as the priority search \kdt data structure. 
    \item We provide fast implementations of our algorithms and perform extensive experimental evaluations showing that our implementations outperform state-of-the-art implementations by up to orders of magnitude.  
\end{list}

Our source code is publicly available at {\small\url{https://github.com/michaelyhuang23/ParCluster}}.

\begin{table*}[t]
\footnotesize
\centering
 \begin{tabular}{ll ccc} 
 \toprule
   & \textbf{Algorithm} & \textbf{worst-case work} & \textbf{average-case work} & \textbf{worst-case span}  \\
 \midrule
  \midrule
    \multirow{5}{*}{Density Computation} & \emph{Original DPC$^*$~\cite{Rodriguez14}} &  $\Theta(n^2)$ & $\Theta(n^2)$  & $\BigO{\log n}$  \\
 & \emph{Exact Baseline DPC$^*$~\cite{Amagata21}} & $\BigO{n(n^{1-\frac{1}{d}}+\varrho_\text{avg})}$ & $\BigO{\min(n(n^{1-\frac{1}{d}}+\varrho_\text{avg}), n\varrho_\text{avg}\log n)}$ & $\BigO{\log n\log\log n}$   \\
 & \emph{R-tree DPC$^*$~\cite{Rasool20}} & $\BigO{n^2}$ & -- & $\BigO{\log^2 n}$   \\
 & \emph{Fenwick DPC (Ours)} & $\BigO{n(n^{1-\frac{1}{d}}+\varrho_\text{avg})}$ & $\BigO{\min(n(n^{1-\frac{1}{d}}+\varrho_\text{avg}), n\varrho_\text{avg}\log n)}$ & $\BigO{\log n\log\log n}$   \\ %
 & \emph{Priority DPC (Ours)} & $\BigO{n(n^{1-\frac{1}{d}}+\varrho_\text{avg})}$ & $\BigO{\min(n(n^{1-\frac{1}{d}}+\varrho_\text{avg}), n\varrho_\text{avg}\log n)}$ & $\BigO{\log n\log\log n}$   \\
 \midrule
 \multirow{5}{*}{Dependent Point Finding} & \emph{Original DPC$^*$~\cite{Rodriguez14}} & $\Theta(n^2)$ & $\Theta(n^2)$  & $\BigO{1}$  \\
& \emph{Exact Baseline DPC$^*$~\cite{Amagata21}} &  $\BigO{n^2}$ & $\BigO{n\log n}$ & $\BigO{n\log n}$  \\
& \emph{R-tree DPC$^*$~\cite{Rasool20}} & $\BigO{n^2}$ & -- & $\BigO{\log^2 n}$  \\ %
& \emph{Fenwick DPC (Ours)} & $\BigO{n^2}$ & $\BigO{n\log^2 n}$ & $\BigO{\log n\log\log n}$  \\
&  \emph{Priority DPC (Ours)} &  $\BigO{n^2}$ & $\BigO{n\log n}$ & $\BigO{\log n\log\log n}$  \\
 \bottomrule
\end{tabular}
\caption{Average-case work and worst-case work and span of DPC algorithms. The span for the algorithms marked with $^*$ is the span of a trivial parallelization of the algorithm. $\varrho_\text{avg}$ is the average density, where the density value of each point equals the number of points around it within a hyper-cubical region with a side length of $2d_\text{cut}$. This is different from the $\rho$ density parameter for DPC, because it is computed as the number of points around it within a hyperball with radius $d_\text{cut}$. %
"--" indicates that we were unable to find a bound with a proof.
}
\label{tbl:complexity}
\end{table*}

\section{Related Work}\label{sec:related}
\subsection{\kdt{s} and $K$-nearest neighbor queries}

In this work, we use \kdt{s} and a variant of it for $K$-nearest neighbor query and range search. There are also variants of \kdt{s} that are specialized for other tasks. \citet{Maneewongvatana99} proved that a \kdt that adopts a sliding mid-point space partitioning scheme only visits $\BigO{K}$ cells,
in the worst case; however, their \kdt does not have a bounded height and therefore does not have a $\BigO{\log n}$ average-case query complexity. \citet{Wald05} proposed implicit \kdt{s}, which define the partitioning of space using a recursive splitting-function and is applied in ray tracing. \citet{Robinson81} proposed the K-D-B-tree, which is used to organize large point sets stored in secondary memory. \citet{Gro07} proposed the min-max \kdt, which is designed for storing points with an extra attribute value. Each node of the min-max \kdt records the minimum and maximum attribute value amongst all points stored under the subtree of that node, which can be used to prune searches~\cite{Wald05, Gro07}. Our proposed priority search \kdt is an optimized variant of a max \kdt. It can also be viewed as a generalization of the priority search tree data structure~\cite{McCreight85} to higher dimensions.

\subsection{Density Peaks Clustering (DPC)}
 Many variants of the standard DPC~\cite{Rodriguez14} have been developed~\cite{Chen19,Mehmood17,Wang16,Yuan20,Du21,Jiang19,Xu21}, and there has also been a line of work focused on improving the computational efficiency of the standard DPC algorithm. The naive DPC algorithm takes $\Theta(n^2)$ work, and can be implemented to take $\BigO{\log n}$ span for density computation and $\BigO{1}$ span for depending point finding.\footnote{Note that although the naive DPC algorithm achieves better span complexity than subsequent DPC algorithms, its work complexity is larger.}
 Bai et al.~\cite{Bai17} utilized $k$-means clustering as a preprocessing step of DPC to prune the number of points needed to be traversed to find a point's density and dependent point. Gong et al.~\cite{Gong16} parallelized DPC in a distributed setting and employed Voronoi diagrams to improve its efficiency. \citet{liu2023improving} proposed a DPC algorithm for GPUs. Amagata and Hara~\cite{Amagata21} leveraged the \kdt data structure to improve the efficiency of density computation and dependent point finding. \citet{Rasool20} used an R-tree to optimize the density computation and dependent point searching efficiency. Their algorithm is sequential, but in theory, it could be parallelized by using a parallel version of R-trees~\cite{hoel1993data} and performing queries in parallel. %
We summarize the complexity of DPC algorithms in \Cref{tbl:complexity}. %
\iffull
The average-case work bounds of our algorithms are proved in Appendix~\ref{sec:complexity}--\ref{sec:range-appendix}.
\fi

Some works have relaxed the definition of DPC to develop efficient algorithms for approximate DPC. \citet{Zhang16} proposed LSH-DDP, a parallel  DPC algorithm for distributed memory that first hashes points into buckets, with spatially-close points being hashed into the same bucket. It then approximates the density and dependent point query of a point $x$ by only considering points from the same bucket as $x$. Finally, it applies corrections to the approximations as necessary. \citet{sieranoja2019fast} developed an algorithm that first constructs a $K$-nearest neighbor graph, and then computes approximate density values and dependent points based on the $K$-nearest neighbor graph.
\citet{Amagata21} also proposed a parallel approximate DPC algorithm that constructs a spatial grid on top of the points. Leveraging the grid structure, the algorithm shares density and dependent point computations across all points inside the same grid cell, thus reducing the computational cost.

\subsection{Density-based Clustering Algorithms}

DPC falls under the broad category of density-based clustering algorithms. Density-based clustering algorithms come in different varieties. Some density-based clustering algorithms define the density of a point based on the number of points in its vicinity~\cite{Ester96,Agrawal98,Ankerst99,Januzaj04,Rodriguez14, sddp, Chen19}. Others leverage a grid-based definition~\cite{Wang97,Hinneburg98,Hanmanthu18,Sheikholeslami00}. Some algorithms define density based on a probabilistic density function~\cite{Wang97,Kriegel05,Smiti13}. One popular density-based clustering algorithm is DBSCAN~\cite{Ester96}, which has many derivatives as well~\cite{Ankerst99,Tepwankul10,Gotz15,Borah04,Ertoz03,Campello2015}.

\section{Preliminaries}\label{sec:prelim}
In this section, we provide the definitions and notations used in this paper. We assume that arrays are indexed from $1$ to $n$.

Let $P = \{x_1, x_2, \ldots, x_n\}$ represent a set of $n$ points that we need to cluster. Each point is given in $d$-dimensional coordinate space. We use $x$ to denote a generic point in $\mathbb{R}^d$ and $x_i$ to represent the $i^\text{th}$ point in our point set $P$. Let $D(x_i, x_j)$ denote the distance between point $x_i$ and point $x_j$. For the complexity results of our work to hold, $D$ should be a metric distance~\cite{Friedman77}.

DPC requires three parameters: $d_\text{cut}$, $\rho_\text{min}$, and $\delta_\text{min}$. Intuitively, $d_\text{cut}$ controls how the density is computed; $\rho_\text{min}$ controls the noise level; and $\delta_\text{min}$ controls the granularity of clusters. Below, we formally explain how they are used.

\begin{definition}
    Given a point $x_i\in P$ and a cutoff value $d_\text{cut}$, we define the \defn{density} of $x_i$ to be $\rho(x_i)=\lvert \{x_j \mid x_j\in P \text{ and } D(x_i, x_j)\le d_\text{cut}  \} \rvert$, i.e., the number of points inside a hyperball centered at $x_i$ with radius $d_\text{cut}$.
\end{definition}

\begin{definition}
   Let $P_i = \{x_j \mid x_j\in P \text{ and } \rho(x_j)>\rho(x_i)\}$.
    For a point $x_i \in P$, the \defn{dependent point} of $x_i$ is a point $\lambda(x_i)\in P_i$ such that,
    $
    D(x_i,\lambda(x_i)) \le D(x_i, x_j)\ \forall \ x_j \in P_i %
    $.
\end{definition}
 Given a point $x_i$, we define its \defn{dependent point set} $P_i$ as the set of points with density value higher than $\rho(x_i)$. When $\rho(x_i)=\rho(x_j)$ for some points $x_i$ and $x_j$, the tie is broken lexicographically. The dependent point $\lambda(x_i)$ of point $x_i$ is thus $x_i$'s nearest neighbor in $P_i$. %
 
\begin{definition}
    Let $\delta(x_i) =  D(x_i,\lambda(x_i))$ be the \defn{dependent distance} of $x_i$. If $x_i$ is the point with highest density in $P$, then it does not have a well-defined dependent point. In that case, we let $\delta(x_i) = \infty$. 
\end{definition}
\begin{definition}
A point $x_i\in P$ is considered a \defn{noise point} if $\rho(x_i)< \rho_\text{min}$ for some density cutoff $\rho_\text{min}$.   
\end{definition}
\begin{definition}
$x_i$ is considered a \defn{cluster center} if $\delta(x_i)\ge \delta_\text{min}$ and it is not a noise point. 
\end{definition}

Each cluster center corresponds to a separate cluster. Each non-noise point that is not a cluster center is assigned to be in the same cluster as its dependent point. Noise points do not belong to any cluster.

 Thus, $d_\text{cut}$, $\rho_\text{min}$, and $\delta_\text{min}$ are the three hyper-parameters of DPC. They can be set manually using the visual aid of an intuitive decision graph that plots each point $x_i$'s density value $\rho(x_i)$ against its dependent point distance $\delta(x_i)$~\cite{Rodriguez14}. \citet{Rodriguez14} suggest that $d_\text{cut}$ can chosen such that the average
number of neighbors is between $0.01n$--$0.02n$. There are also automatic parameter tuning methods~\cite{garcia2021methodology, yin2022improved}.

\subsection{Model of Computation}
We use the \defn{work-span model}~\cite{JaJa92,CLRS}, a standard model for analyzing shared-memory parallel algorithms.  
The \defn{work} $T_1$ of an algorithm is the total number of operations executed by the algorithm, and the \defn{span}  $T_\infty$ is the length of the longest sequential dependency chain of the algorithm (it is also the parallel time complexity when there are an infinite number of processors).
We can bound the expected running time of an algorithm on $\mathcal{P}$ processors by $T_1/\mathcal{P} + O(T_\infty)$ using a randomized work-stealing scheduler~\cite{Blumofe1999}.

\subsection{Relevant Techniques}\label{sec:relevant-technique}
Our algorithms make use of the \kdt~\cite{Bentley75} and Fenwick tree~\cite{Fenwick94} data structures. 

\myparagraph{\kdt{s}}
A \defn{\kdt{}}~\cite{Bentley75} is a binary space partitioning tree, where each internal node
contains a splitting hyperplane that partitions the points contained
in the node between its two children. Let the smallest bounding box containing all points in a node be the node's \defn{cell}.
The root node contains all of the points, and the \kdt is constructed by recursing on each of its two children after splitting, until a leaf node is reached. Each node stores the coordinates for its cell, which can be used for pruning searches. The \kdt can be constructed with $\BigO{n\log n}$ work and $\BigO{\log n\log{\log n}}$ span \cite{wang2022esa,Yesantharao22}.
\kdt{s} can answer two types of queries efficiently: finding points inside a radius and 
finding the nearest neighbors of some chosen point.
We call the first type \defn{range query} and the second 
\defn{nearest neighbor query}.

A \kdt $T$ can be incremental, in which case we can insert points into $T$.
Note that an incremental \kdt can be imbalanced and not satisfy complexity results of a normal \kdt. 
We use \algname{build-\kdt}$(P)$ to represent initializing a \kdt from the set of points $P$.

\myparagraph{Range query with \kdt{s}} Let $T.$\algname{query-range}$(x, r)$ denote a range search on $T$, in a spherical region $R$ with radius $r$ centered at a point $x$, and returns the number of points inside the region. 
In a range query, when traversing down the \kdt, we only need to visit a node if its cell intersects with $R$; otherwise the node can be pruned from the search. 
A range query takes $\BigO{n^{1-\frac{1}{d}}+|Q|}$ work on a balanced \kdt with splitting dimension chosen cyclically, where $Q$ is the set of points returned and $d$ is the dimension of the data set~\cite{bentley1979data}. The query takes $\BigO{\log n}$ span by visiting children in parallel.

\myparagraph{Nearest neighbor query with \kdt{s}}
We use $T.$\algname{query-nn}$(x)$ to represent performing a nearest neighbor search on $T$ for the point $x$, which returns the closest neighbor of $x$. 
To compute the nearest neighbor of a point $x$,
 we first traverse down the \kdt to find the leaf node that contains the point $x$. Then, in the backtracking process, we search the sibling subtrees. Let $x$'s distance to the current nearest neighbor candidate of $x$ be represented by $L$. We prune the search of any subtree whose cell is farther than $L$ away from $x$.  \citet{Friedman77} proved that the average-case work complexity of a nearest neighbor search can be bounded by $\BigO{\log n}$ under the assumptions that the density of points in space is locally uniform and the \kdt is split at the widest dimension's median per level.

\myparagraph{Fenwick tree}\label{sec:prelim:fenwick}
The Fenwick tree decomposes a range $[1, n]$ into $n$ sub-ranges such that the $i^\text{th}$ sub-range, represented by $B[i]$, corresponds to the range $[i-\text{LSB}(i)+1, i]$, where $\text{LSB}(i)$ represents the least significant bit of integer $i$ and $\sum_{i=0}^n |B[i]| = \BigO{n\log n}$ \cite{Fenwick94}. The key property of a Fenwick tree is that each prefix range $[1, i]$ can be decomposed into $\BigO{\log n}$ disjoint sub-ranges; we represent the set of these sub-ranges by $S[i]$. In other words, 
$\bigcup_{j\in S[i]} B[j] = [1, i]$.
Each $S[i]$ can be built iteratively in $\BigO{\log n}$ work, and we can access a partition of the range $[1,i]$ in $\BigO{\log n}$ work using the indices stored in $S[i]$. 

\myparagraph{Union-find}\label{sec:prelim:union}
The union-find data structure maintains the set membership of elements, and allows for merging of these sets. Initially, each element is in its own singleton set. A \algname{union}($a,b$) operation merges $a$ and $b$ into the same set.
We use a lock-free concurrent union-find~\cite{Jayanti21}, where performing $m$ unions on a union-find data structure with $n$ elements takes $\BigO{m\left( \log(\frac{n}{m}+1) + \alpha(n,n) \right)}$ work and $\BigO{\log n}$ span ($\alpha$ denotes the inverse Ackermann function).

\myparagraph{Other parallel primitives}
\algname{write-min}$(loc, val)$ 
is a priority concurrent write that takes as input two
arguments, where the first argument is the location to write to and the
second argument is the value to write; on concurrent writes,
the smallest value is written~\cite{ShunBFG2013}. %
We assume that \algname{write-min} takes $O(1)$ work and span.
\algname{radix-sort$(A)$} takes a sequence of elements $A$ of size $n$, with an ordering key defined for each element. It sorts them in parallel according to the ordering of the elements' keys. Radix sort takes $\BigO{n}$ work and $\BigO{\log n}$ span with high probability (\whp{})\footnote{We say $O(f(n))$ \emph{with high probability (\whp{})} to indicate
$O(cf(n))$ with probability at least $1-n^{-c}$ for $c \geq 1$,
where $n$ is the input size.} given that the range of the keys is bounded by $\BigO{n\log^{O(1)}n}$~\cite{Rajasekaran89}.

\section{Priority Search \kdt-based Dependent Point Finding}\label{sec:priority-kd-tree}
We present our first algorithm for solving the dependent point finding task using our new variant of \kdt{s}.  

\subsection{Sequential Dependent Point Finding}\label{sec:incomplete-kd-tree}
To warm up, we first introduce a sequential incomplete \kdt based algorithm for finding dependent points. 
This algorithm is an improvement over \citet{Amagata21}'s dependent point finding algorithm. Their algorithm uses an incremental \kdt, which incurs an expensive cost for inserting points.\footnote{Although \citet{Amagata21}'s exact DPC algorithm has parallel components, their dependent point finding step is sequential.} In \citet{Amagata21}'s algorithm, points are sorted in reverse order of density, and inserted to the tree one by one in order via top-down traversals of the tree. Each point queries its nearest neighbor in the tree, before being inserted into the tree itself. Since points are inserted in reverse order of density, the nearest neighbor of a point $x$ as returned by the incremental \kdt must have a higher density value than $x$ and be its dependent point~\cite{Amagata21}.

\begin{figure}[t]%
\vspace{-5pt}
    \centering
    \includegraphics[width=0.3\columnwidth]{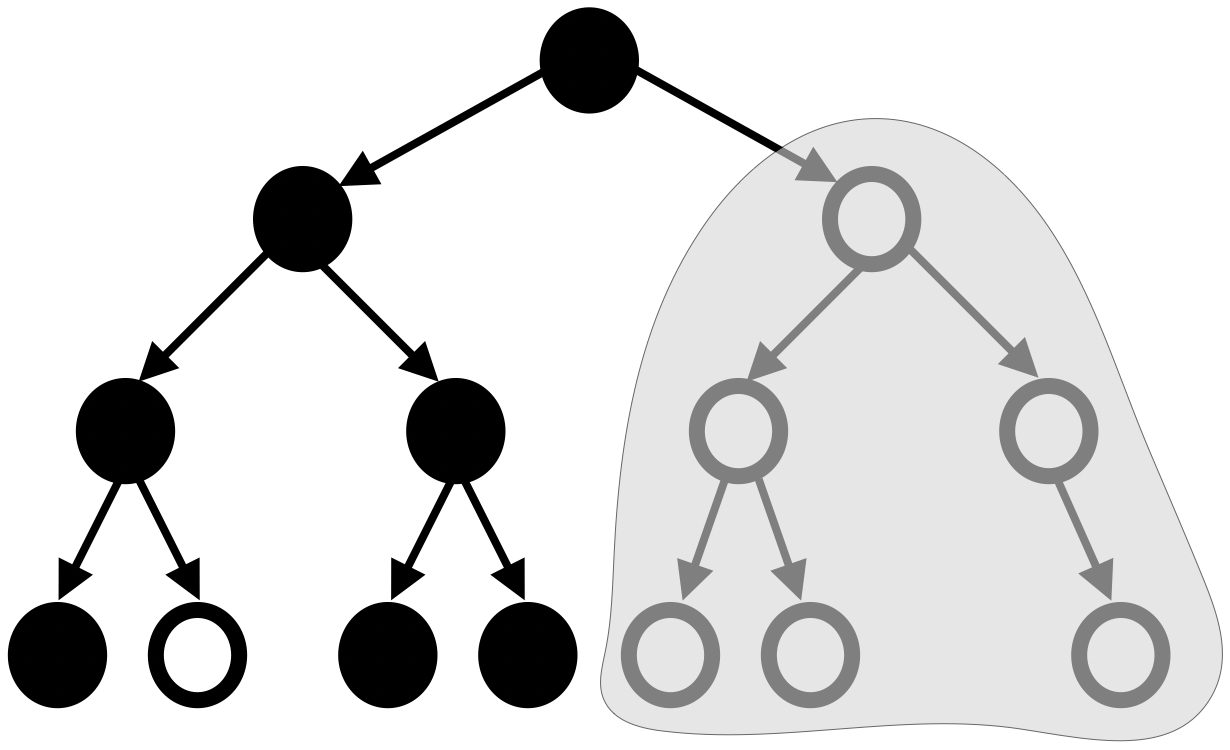}
    \caption{An example of an incomplete \kdt. A node is unfilled if its subtree does not contain any active point; otherwise it is filled. During a nearest neighbor search, the entire grayed out subtree can be pruned because it contains no active point.}\label{fig:incomplete-kd-tree}
\end{figure}

We propose to use an incomplete \kdt in place of an incremental \kdt, and take advantage of the fact that we know all the points to insert. Instead of inserting points into the incremental \kdt, we utilize a lazy insertion strategy: a balanced \kdt is constructed with all points in $P$, but all points are marked as inactive initially. We use a boolean variable $\text{isActive}_i$ (initialized to \emph{false}) to track if the $i^\text{th}$ subtree contains an active point. When we insert a point into the \kdt, we simply activate the point and set $\text{isActive}$ to $\emph{true}$ for all of its ancestors in the tree by a bottom-up traversal from the leaf node containing the inserted point.
When traversing the \kdt to query for the nearest neighbor, we can prune a subtree $i$ if its $\text{isActive}_i$ value is \emph{false}. An example of incomplete \kdt is given in Figure \ref{fig:incomplete-kd-tree}.

Our new method has two advantages. First, the incremental \kdt can be imbalanced and make querying slower while our \kdt is always balanced since its structure is not modified after construction. This is especially important when we perform queries in parallel, because the span of the query is the same as the depth of the tree. Second, traversing down the \kdt to insert a point requires computing which child the point belongs to, but our method only needs to follow the parent pointers starting at a leaf node to traverse up the tree without requiring any other computation.

However, this method still has high span, because the points are inserted one by one. The span of each nearest neighbor search is $\BigO{\log n}$, and there are $\BigO{n}$ nearest neighbor queries, so the span is $\BigO{n \log n}$. \citet{Amagata21}'s algorithm can also be modified to achieve $\BigO{n \log n}$ span by replacing their sequential nearest neighbor queries with parallel nearest neighbor queries. In the rest of the section, we will see that we can use a priority search \kdt to find the dependent points of all points in parallel.

\subsection{Priority Search \kdt}\label{sec:priority-search-kdt}
\subsubsection{Priority Search \kdt Definition}
\begin{figure}[t]%
\vspace{-3pt}
    \centering
    \includegraphics[trim={25cm 11cm 23cm 10cm},clip, width=0.4\columnwidth]{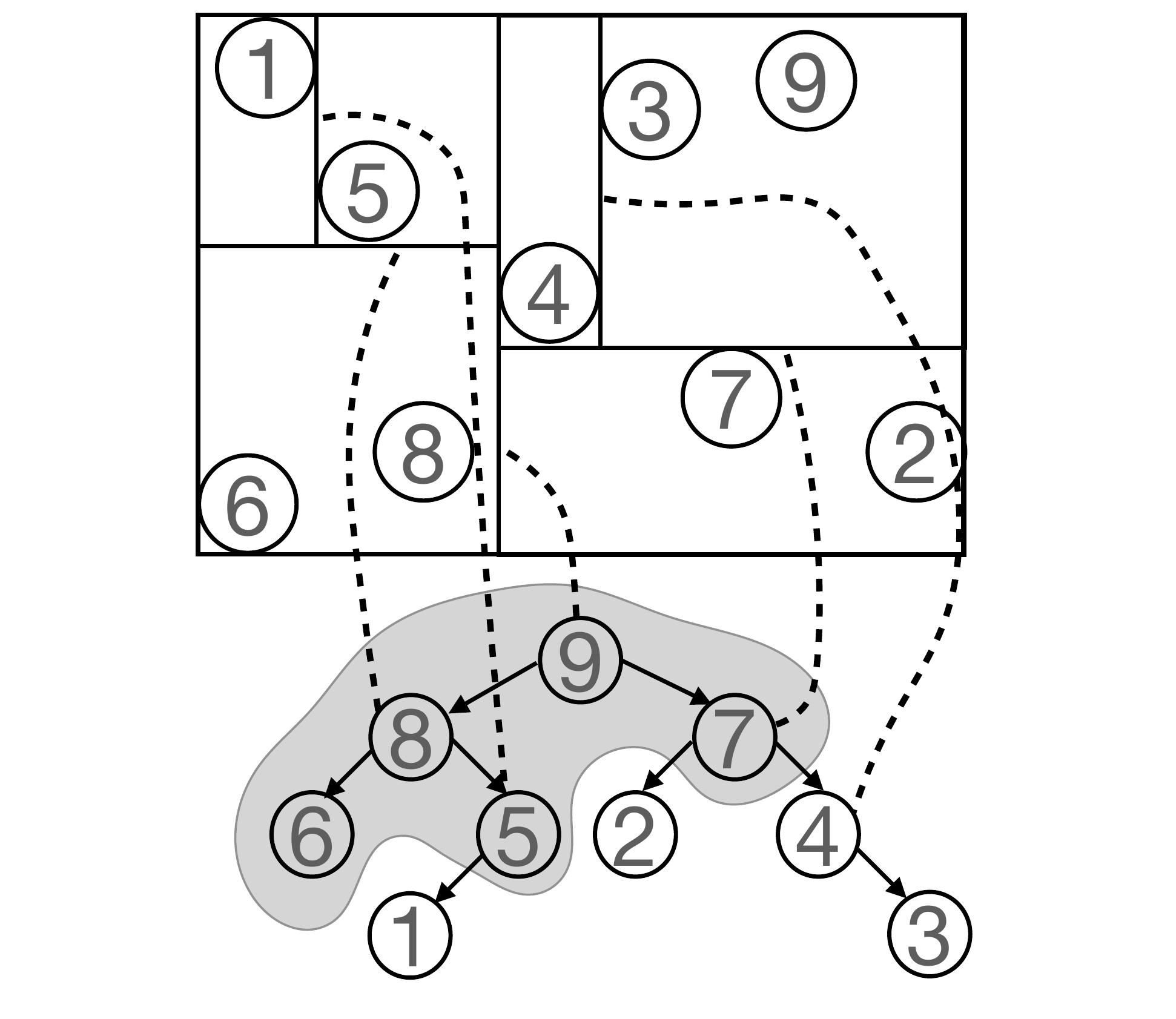}
    \caption{An example of a priority search \kdt. Each point is labeled with its priority value $\gamma$, which is an integer from $1$ to $9$ in this example. Each node of the priority search \kdt stores the point with the highest $\gamma$ within the region of the cell of the node; the number inside the circle of the node represents the node's $\gamma$ value. The dotted lines on the graph connects each node with the splitting hyperplane of that node. The grayed area represent a subgraph $T_q$ comprising all nodes with $\gamma > 4$. Because the $\gamma$ values of a priority \kdt satisfy the heap property, $T_q$ is always an upper portion of the priority search \kdt.}\label{fig:priority-kdt}
\end{figure}

To parallelize the dependent point finding routine described in Section~\ref{sec:incomplete-kd-tree}, we first introduce a parallel analogue of the incomplete \kdt---a \defn{priority search \kdt}---and describe its general properties. 
A priority search \kdt is designed to store a set of points $P=\{x_1, x_2, \ldots, x_i, \ldots, x_n\}$, such that each point $x_i\in \mathbb{R}^d$ is associated with a priority value $\gamma_i$. In our case, $\gamma_i$ is the density. 
Similar to a normal \kdt, each node in the priority search \kdt corresponds to a set of points and a partition of space. Additionally, each node in a priority search \kdt stores the point with the highest $\gamma$ value among all points in its point set; this $\gamma$ value is referred to as the $\gamma$ value of the node. The remaining points are split evenly between the children of the node along a hyperplane perpendicular to the longest side of the cell of that node.
An example of a priority search \kdt is shown in Figure~\ref{fig:priority-kdt}. A priority search \kdt is structurally similar to a max \kdt~\cite{Gro07}, which records only the maximum priority value at each node. However, in a max \kdt, the point with the maximum priority value is stored at a leaf in either the left or the right subtree of that node, instead of directly at that node like our priority search \kdt. 
A priority search \kdt is advantageous in that a meaningful priority range query complexity bound can be established for it, but not for a max \kdt because each cell in a max \kdt is not uniquely associated
with a point. 
\iffull
We give more details in Section~\ref{sec:priority-range-query}. 
\fi

Priority search \kdt{s} can be constructed similarly to a normal \kdt; the only extra step is finding the point with highest priority value at each node by scanning all of its descendant nodes. The cost of this extra step is subsumed by the cost of splitting at each node. Construction takes $\BigO{n\log n}$ work and $\BigO{\log n\log{\log n}}$ span. The data structure takes $\BigO{n}$ space like a normal \kdt, because only $\BigO{1}$ extra information is stored at each node.%

\subsubsection{Priority Nearest Neighbor Query}\label{sec:priority-search-kdt-knn}

Let the \defn{priority nearest neighbor} be the nearest neighbor of a point $x_i$ with higher priority. It is easy to see that the priority nearest neighbor query problem is equivalent to the problem of finding dependent points if we set the priority value $\gamma_i$ for a point $x_i$ to be the density value $\rho(x_i)$. A more formal definition is given below.
\begin{definition}
    Given a generic query point $x_q\in \mathbb{R}^d$, a distance measure $D$, and a point set $P\subseteq \mathbb{R}^d$, we define $P_q = \{ x_i \mid x_i \in P\ \text{and}\ \lambda_i > \lambda_q\}$. Then, the priority nearest neighbor of $x_q$ is the point in $P_q$ that is closest to $x_q$ as measured by $D$.
\end{definition}

Priority nearest neighbor queries can be solved by querying a priority search \kdt following a similar procedure as a normal nearest neighbor query, with the exception that all subtrees with priority value $\le \lambda_q$ are pruned from the search. 
Consider a particular query, with a threshold priority value of $\lambda_q$. Let $T$ denote a priority search \kdt.
Let $T_q\subseteq T$ represent the set of nodes with priority value $>\gamma_q$. Because of the structure of the priority search \kdt, $T_q$ must be a connected subgraph of $T$.\footnote{If two tree nodes $a$ and $b$ are both in $T_q$ but not connected, then some ancestor of either $a$ or $b$ is not in $T_q$ and has a priority value less than $\gamma_q$. However, this is not possible because the ancestors can only have higher priority values.} 
A priority nearest neighbor search on $T$ is thus equivalent to a normal nearest neighbor search on an incomplete \kdt $T$ with $T_q$ forming its active portion. Thus, similar to the complexity result on an incomplete \kdt, a priority nearest neighbor query on a priority search \kdt takes $\BigO{\log n}$ average-case work with some assumptions, $\BigO{n}$ worst-case work, and $\BigO{\log n}$ worst-case span. 
Similar assumptions are made in the analysis of the original \kdt~\cite{Friedman77}. 
For average-case analysis, we assume that the points in $P_q$ are sampled from $\mathbb{R}^d$ according to some probability density function $\mu_q$, and that the number of points is sufficiently large such that $\mu_q$ can be considered locally uniform. 
\iffull
We provide a proof of the bounds in Section \ref{sec:complexity}.
\fi

\subsection{Parallel Dependent Point Finding with Priority Search \kdt}\label{sec:priority-dep}
We now apply the priority search \kdt to solve the dependent point finding task. 
The recipe for finding dependent points using a priority search \kdt is given in Algorithm \ref{alg:priority}, where $P$ is the input data set with size $n$ and $\rho$ is an array containing the density of points in $P$.
On Line~\ref{alg:priority:buildtree}, we construct the priority search \kdt in $\BigO{n\log n }$ work and $\BigO{\log n \log\log n }$ span. %
On Lines~\ref{alg:priority:query-for}--\ref{alg:priority:dcomp}, we compute the dependent point for each non-noise point in parallel.
Let the work and span of each dependent point search operation be $\mathcal{W}$ and $\mathcal{S}$, respectively, on the priority search \kdt. The work and span of this step is then $\BigO{n\mathcal{W} + n\log n}$ and $\BigO{\mathcal{S} + \log n \log\log n }$, respectively. In the worst case, $\mathcal{W}=\BigO{n}$ and $\mathcal{S}=\BigO{\log n}$. Under the assumptions stated earlier, $\mathcal{W}=\BigO{\log n}$ in the average case.

\begin{theorem}
Constructing a priority search \kdt and finding all dependent points can be done in $\BigO{n\log n}$ average-case work, $\BigO{n^2}$ worst-case work, and $\BigO{\log n\log{\log n}}$ worst-case span. The space usage is $\BigO{n}$.
\end{theorem}

\begin{algorithm}[!t]
\vspace{-3pt}
 \footnotesize
\caption{Parallel dependent point finding with a priority search \kdt}
 \begin{algorithmic}[1]
 \Procedure{priority-dependent-point}{$P$, $\rho$} 
 \State $T\leftarrow $ \algname{build-priority-search-kd-tree}$(P, \rho)$ \label{alg:priority:buildtree}
 \ParFor{all $x_{i}$ in $P$ with $\rho(x_i) \geq \rho_\text{min}$} \label{alg:priority:query-for}
  \State $\lambda(x_{i})\leftarrow T$.\algname{query-priority-nn}$(x_{i})$ \label{alg:priority:dcomp}
 \EndParFor
 \State \Return $\lambda$
 \EndProcedure
 
 \end{algorithmic}
 \label{alg:priority}
\end{algorithm}

\section{Fenwick Tree-based Parallel Dependent Point Finding}\label{sec:fenwick-kd-tree}
We introduce another parallel algorithm for solving the dependent point finding task, which is based on Fenwick trees. This algorithm has the same overall span  as the priority search \kdt algorithm, but has a higher work complexity. However, the average-case work complexity analysis does not make assumptions about the density of $P_q$. It only assumes that $P$'s underlying probability density function is locally uniform (the standard assumption for a \kdt's $\BigO{\log n}$ average-case nearest neighbor query complexity to hold). 

This algorithm can be summarized as follows. We first construct an array $\overline{P}$ of points in $P$ sorted by descending order of their density values. Assume that $\overline{P}$ is indexed from $1$ to $n$. 
Then, we construct a Fenwick tree decomposition of the range $[1, n]$. 
$B[i]$ contains the \kdt that has points in $\overline{P}$ with indices $[i-\text{LSB}(i)+1, i]$.
Recall that $S[i]$ represents a decomposition of the range $[1,i]$ into sub-ranges that are inside $B$. To perform  the dependent point query for the $i^\text{th}$ point in array $\overline{P}$, we simply need to search through every \kdt that corresponds to a sub-range in $S[i-1]$. These queries can be done in parallel, thus giving a low span complexity.  

\begin{algorithm}[!t]
 \footnotesize
\caption{Parallel dependent point finding with a Fenwick tree}
 \begin{algorithmic}[1]

  \Procedure{fenwick-query}{$B, i, x$} 
 \State ($\delta, \lambda')\leftarrow (\infty, \infty)$
 \State Build $S[i]$ \Comment{build a list of indices whose corresponding sub-ranges span the range $[1, i]$} \label{alg:fenwick:builds}
 \ParFor{all $j$ in $S[i]$}\label{alg:fenwick:qstart}
  \State $y \leftarrow B[j].$\algname{query-nn}$(x)$ \label{alg:fenwick:kd-tree-query}
    \State \algname{write-min}$((\delta, \lambda'), (\text{dist}(x, y), y))$ \label{alg:fenwick:atomic-update}
  \EndParFor\label{alg:fenwick:qend}
  \State \Return $\lambda'$
 \EndProcedure

 \Procedure{fenwick-dependent-point}{$P$, $\rho$} 
 \State $\overline{P}\leftarrow$ \algname{radix-sort}$(P)$ \Comment{sorting in descending order of density}\label{alg:fenwick:rsort}
  \State Initialize $B$ as an array of length $n$ \label{alg:fenwick:itrees}
  \State Initialize $\lambda$ as an array of length $n$, with values all being  $\infty$  \label{alg:fenwick:lambda} %
 \ParFor{$i=1$ to $n$} \label{alg:fenwick:buildtrees1}
  \State $B[i]\leftarrow$ \algname{build-\kdt}($\overline{P}[i - \text{LSB}(i) + 1, i]$) \label{alg:fenwick:buildtrees2}
 \EndParFor
 \ParFor{$x_{i}$ in $\overline{P}[2:n]$ with $\rho(x_i) \geq \rho_\text{min}$} \label{alg:fenwick:query-forr}
  \State $\lambda(x_{i})\leftarrow $ \algname{fenwick-query}$(B, i-1, x_{i})$ \label{alg:fenwick:dcomp} %
 \EndParFor
 \State \Return $\lambda$ 
 \EndProcedure
 \end{algorithmic}
 \label{alg:fenwick}
\end{algorithm}

We provide the pseudocode for the algorithm in Algorithm \ref{alg:fenwick}. The main procedure is \algname{fenwick-dependent-point}$(P, \rho)$, which takes as input an array of points $P$ and an array  $\rho$ containing the computed densities of the points. %
On Lines~\ref{alg:fenwick:rsort}--\ref{alg:fenwick:lambda}, we first initialize the $\overline{P}$ array containing points sorted in descending order of their density values, an array $B$ 
to store the $n$ \kdt{s} in the algorithm, and an array $\lambda$ to store the dependent points. On Lines~\ref{alg:fenwick:buildtrees1}--\ref{alg:fenwick:buildtrees2}, we construct the $n$ \kdt{s}; the $i^\text{th}$ \kdt $B[i]$ is constructed from the range of points $\overline{P}[i-\text{LSB}(i)+1, i]$. Finally, on Lines~\ref{alg:fenwick:query-forr}--\ref{alg:fenwick:dcomp}, we perform \algname{fenwick-query} for non-noise 
 points in parallel to find the dependent point for all points. We do not need to find the dependent point for $x_1$ as it is the point with highest density.

Now, we will explain procedure \algname{fenwick-query}, which takes as input an array of \kdt{s} $B$, an index $i$, and a point $x$; \algname{fenwick-query} performs a nearest neighbor query for point $x$ among the points $x_1, x_2, \ldots, x_i$. On Line \ref{alg:fenwick:builds}, we construct a set $S[i]$ for the input index $i$, which contains the indices of the Fenwick tree sub-ranges that form a partition of $[1, i]$, as described in Section \ref{sec:prelim:fenwick}. Each of these sub-ranges corresponds to a \kdt; we perform nearest neighbor queries \algname{query-nn} on all of these \kdt{s} on Line \ref{alg:fenwick:kd-tree-query}. Let $\lambda'$ signify the current dependent point of point $x$ and $\delta$ the distance between $x$ and $\lambda'$. On Line \ref{alg:fenwick:atomic-update}, the dependent point with the smallest distance to $x$ (breaking ties using the point ID) is computed using the concurrent \algname{write-min} function. The current $\lambda'$ is replaced by a newly found nearest neighbor if the newly found nearest neighbor is closer to $x$ than $\lambda'$ is.

\myparagraph{Analysis}
We first analyze the complexity of the \algname{fenwick-query} subroutine. We show that it takes $\BigO{\log^2 n}$ average-case work and $\BigO{n}$ worst-case work. The construction of $S$ on Line~\ref{alg:fenwick:builds} takes $\BigO{\log n}$ work~\cite{Fenwick94}. Each call of \algname{query-nn} on Line~\ref{alg:fenwick:kd-tree-query} takes $\BigO{\log n}$ average-case work~\cite{Friedman77}, which sums to $\BigO{\log^2 n}$ work across all iterations of the parallel for-loop on Line~\ref{alg:fenwick:qstart}. In the worst case, however, each \kdt nearest neighbor query takes time linear in the number of points in the \kdt~\cite{Friedman77}, meaning that Line~\ref{alg:fenwick:kd-tree-query} takes $\BigO{|B[j]|}$ work for the $j^\text{th}$ \kdt, $T_j$. Across all iterations of the parallel for loop, the worst-case work complexity is $\BigO{\sum_{j\in S[i]} |B[j]|} = \BigO{i} = \BigO{n}$. 

In terms of span, \algname{fenwick-query} has a worst-case span of $\BigO{\log n}$. The construction of $S[i]$ takes $\BigO{\log n}$ span. The \algname{write-min} operation on Line~\ref{alg:fenwick:atomic-update} takes $\BigO{1}$ span.
The nearest neighbor query on Line~\ref{alg:fenwick:kd-tree-query} takes a worst-case span of $\BigO{\log n}$ %
since each branch of the \kdt can be searched in parallel and \kdt{s} have $O(\log n)$ depth~\cite{Friedman77}. 
Since all nearest neighbor queries are executed in parallel, the span for the entire \algname{fenwick-query} subroutine is $\BigO{\log n}$.

Now, we examine the main process \algname{fenwick-dependent-point}. We show that its average-case work complexity is $\BigO{n\log^2 n}$ and its worst-case work complexity is $\BigO{n^2}$. Line \ref{alg:fenwick:rsort} takes $\BigO{n}$ work since the keys of the sort---the $\rho$ values---are bounded in value by $\BigO{n}$. On Lines~\ref{alg:fenwick:buildtrees1}--\ref{alg:fenwick:buildtrees2}, constructing the $i^\text{th}$ \kdt takes $\BigO{|B_i|\log |B_i|}$ work. Therefore, constructing all \kdt{s} takes $\BigO{\sum_{i=1}^n |B_i|\log |B_i|}=\BigO{n\log^2 n}$ work. Finally, the \algname{fenwick-query} operations performed in the parallel for-loop on Lines~\ref{alg:fenwick:query-forr}--\ref{alg:fenwick:dcomp} take $\BigO{n\log^2 n}$ average-case work and $\BigO{n^2}$ worst-case work. Overall, \algname{fenwick-dependent-point} takes $\BigO{n\log^2 n}$  average-case work and $\BigO{n^2}$ worst-case work.

Next, we analyze the span of \algname{fenwick-dependent-point}. The radix sort on Line \ref{alg:fenwick:rsort} takes $\BigO{\log n}$ span \whp{}~\cite{Rajasekaran89}. Each \algname{build-\kdt} operation on Line \ref{alg:fenwick:buildtrees2} takes span $\BigO{\log n\log\log n}$. Finally, each call to the subroutine \algname{fenwick-query} on Line \ref{alg:fenwick:dcomp} takes $\BigO{\log n}$ worst-case span. Thus, the overall span of \algname{fenwick-dependent-point} is $\BigO{\log n\log\log n}$ in the worst case. 

Finally, we consider the space usage of our algorithm. The $i^\text{th}$ \kdt, $T_i$, takes space $\BigO{|B_i|}$. Thus, the overall space usage is $\BigO{\sum_{i=1}^n |B_i|}=\BigO{n\log n}$.

 \begin{theorem}
Constructing a Fenwick tree and finding all dependent points can be done in $\BigO{n\log^2 n}$ average-case work, $\BigO{n^2}$ worst-case work, and $\BigO{\log n\log{\log n}}$ worst-case span. The space usage is $\BigO{n\log n}$.
 \end{theorem}
\section{Optimization of Other Steps}\label{sec:par-other}
\subsection{Optimizing Density Computation (Step~\ref{item:density})}\label{sec:optim-density}
\iffull
In Appendix~\ref{sec:range-appendix}, we show that the DPC algorithm's density computation takes $\BigO{\min(\BigO{n(n^{1-\frac{1}{d}}+\varrho_\text{avg})}, n\rho_\text{avg}\log n)}$ average-case work in theory, where $\rho_\text{avg}$ is the average number of points inside the hyper-ball query region with radius $d_\text{cut}$.
\fi
We now discuss an optimization that improves the performance of density computation in practice.
Let $R$ denote the spherical region with radius $r$ and centered at $x^\text{center}$.
Since we only want the count of points in $R$, we do not have to visit every point.
If a cell corresponding to a subtree is contained completely inside $R$, then we can simply add the number of points inside that cell to the count and prune the subtree from the rest of the traversal, instead of visiting every point. We can check whether a hyper-rectangular region in coordinate space is contained inside a sphere $R$ by finding the corner of the region $x_\text{far}$
that is farthest from the center of $R$ and checking if $x_\text{far}$ is enclosed in $R$.

\begin{algorithm}[t!]
 \footnotesize
\caption{Single-linkage clustering with parallel union-find }
 \begin{algorithmic}[1]

 \Procedure{single-linkage-cluster}{$P$, $\lambda$, $\delta_{\min}$, $\rho_{\min}$} 
  \ParFor{all $x_{i}$ in $P$} \label{alg:single-linkage-cluster:query-for}
  \If{$\lambda(x_i) \not= \infty$} $\delta(x_i)\leftarrow \text{dist}(x_i, \lambda(x_i))$\EndIf  \label{alg:single-linkage-cluster:store}%
 \EndParFor
 \State Initialize $F$ to be an empty parallel union-find data structure \label{alg:single-linkage-cluster:init}
 \ParFor{all $x_i$ in $P$}
    \If{$\delta(x_i) < \delta_\text{min}$ or $\rho(x_i)<\rho_\text{min}$} \Comment{check if $x_i$'s dependent distance is $<$ threshold}
        \State $F.$\algname{union}$(x_i, \lambda(x_i))$ \label{alg:priority:unionize}
    \EndIf
 \EndParFor
 \State \Return $F.\text{cluster-labels}$
 
 \EndProcedure
 
 \end{algorithmic}
 \label{alg:single-linkage-cluster}
\end{algorithm}

\subsection{Optimizing Single-Linkage Clustering (Step~\ref{item:single-linkage})}
In this subsection, we optimize Step~\ref{item:single-linkage} of DPC.
We use a lock-free parallel union-find data structure~\cite{Jayanti21} to solve single-linkage clustering, thus cutting down the $\BigO{n}$ span complexity from \citet{Amagata21}'s algorithm to $\BigO{\log n}$. 
Our algorithm is shown in Algorithm \ref{alg:single-linkage-cluster}. It takes an array of points $P$, an array of their dependent points $\lambda$, and the parameters $\delta_{\min}$ and $\rho_{\min}$. On Lines~\ref{alg:single-linkage-cluster:query-for}--\ref{alg:single-linkage-cluster:store}, we compute the dependent distance of all points in parallel, which takes $O(n)$ work and $O(1)$ span. 
On Lines~\ref{alg:single-linkage-cluster:init}--\ref{alg:priority:unionize}, we use union-find to cluster points with their dependent points if their dependent distance is less than $\delta_{\min}$ or if their density is less than $\rho_{\min}$.
The initialization on Line \ref{alg:single-linkage-cluster:init} takes $\BigO{n}$ work and $\BigO{1}$ span. 
On Line~\ref{alg:priority:unionize}, performing $O(n)$ unions on a union-find data structure with $n$ elements takes $\BigO{n\alpha(n, n)}$ work~\cite{Jayanti21} and $\BigO{\log n}$ span, and this is also the overall work and span.

\section{Experiments}\label{sec:experiment}
Finally, in this section, we perform experiments on the efficiency of our dependent point finding algorithms as well as our proposed optimizations to density computation. 
\subsection{Experiment Setup}

We run experiments on both real-world and synthetic data sets. The real-world data sets that we use are \emph{GeoLife}~\cite{Zheng08}, \emph{PAMAP2}~\cite{Reiss12}, \emph{Sensor}~\cite{Burgues18,Burgues18-2}, \emph{HT}~\cite{Huerta16}, and \emph{Gowalla}~\cite{cho2011friendship, snapnets}. The synthetic data sets that we use are produced by the \emph{simden} and \emph{varden} random walk based generators by \citet{Gan17}. \emph{Simden} generates multiple clusters of points with similar density while \emph{varden} produces multiple clusters with varying density. We also use synthetic data sets generated by a \emph{uniform} sampler. In addition, we use a synthetic data set \emph{Query}~\cite{UCI1URL, anagnostopoulos2018scalable}. Details of these data sets are listed in Table \ref{tab:datasets} along with the hyper-parameters that we use for each data set. The $d_\text{cut}$ hyper-parameter is selected such that the computed density values based on the chosen $d_\text{cut}$ value is nonzero but significantly smaller than the size of the data set. The $\rho_\text{min}$ and $\delta_\text{min}$ values are selected such that the total number of clusters produced by the DPC algorithm is relatively small. We use Euclidean distances in our experiments.

\begin{table}[!t]
\footnotesize
\centering
\vspace{-5pt}
 \begin{tabular}{l c c c c c} 
 \toprule
 Name & $n$ & $d$ & $d_\text{cut}$ & $\rho_\text{min}$ & $\delta_\text{min}$ \\ %
 \midrule
 \emph{uniform} & $10^3$ to $10^7$ & 2 & 30 & 0 & 100 \\
 \emph{simden} & $10^3$ to $10^7$ & 2 & 30 & 0 & 100 \\
 \emph{varden} & $10^3$ to $10^7$ & 2 & 30 & 0 & 100 \\
 \emph{GeoLife} &  24876978 & 3 & 1 & 1000 & 10 \\
 \emph{PAMAP2} & 259803 & 4 & 0.02 & 20 & 0.2 \\
 \emph{Sensor} &  3843160 & 5 & 0.2 & 5 & 2 \\
 \emph{HT} & 928991 & 8 & 0.5 & 30 & 10 \\
 \emph{Query} & 50000 & 3 & 0.01 & 0 & 0.05  \\
 \emph{Gowalla} & 1256248 & 2 & 0.03 & 0  & 40 \\
 \bottomrule
\end{tabular}
\caption{\label{tab:datasets}The real world data sets used in our experiments, along with their sizes ($n$), their dimensionality ($d$), and the clustering hyper-parameters that we select for them. 
The numbers for Query and Gowalla are after de-duplication.}
\end{table}

\myparagraph{Computational Environment}
We use \textit{c2-standard-60} instances on the Google Cloud Platform. These are 30-core machines with two-way hyper-threading with Intel 3.1 GHz Cascade Lake processors that can reach a max turbo clock-speed of 3.8 GHz. 

\myparagraph{Algorithms}
We implement our algorithms using the  ParlayLib~\cite{Blelloch20} and ParGeo~\cite{wang2022esa} libraries. We use C++ for all implementations, and the gcc compiler with the -O3 flag to compile the code.
We evaluate the following algorithms. 
\begin{list}{\textbullet}{%
    \setlength{\leftmargin}{0.5em}
    \setlength{\itemindent}{0em}
    \setlength{\labelwidth}{\itemindent}
    \setlength{\labelsep}{0.5em}
    \setlength{\listparindent}{1em}
    \setlength{\itemsep}{0em}
    \setlength{\parsep}{0em}
    \setlength{\topsep}{0em}
    \setlength{\partopsep}{0em}
}
    \item \algname{dpc-exact-baseline}: \citet{Amagata21}'s state-of-the-art implementation of the partially parallel exact DPC algorithm, which computes the densities in parallel.
    \item \algname{dpc-approx-baseline}: \citet{Amagata21}'s fastest parallel approximate DPC algorithm.
    \item \algname{dpc-incomplete}: our partially parallel DPC algorithm that uses the incomplete \kdt based dependent point finding algorithm in Section \ref{sec:incomplete-kd-tree} along with the optimizations in Section~\ref{sec:par-other}.
    \item \algname{dpc-priority}: our parallel DPC algorithm that uses the priority search \kdt based dependent point finding algorithm in Section \ref{sec:priority-dep} along with the optimizations in Section~\ref{sec:par-other}.
    \item \algname{dpc-Fenwick}: our parallel DPC algorithm that uses the Fenwick tree based dependent point finding algorithm in Section \ref{sec:fenwick-kd-tree} along with the optimizations  in Section~\ref{sec:par-other}. 
\end{list}

We also compare with \citet{Rasool20}'s sequential DPC algorithm. We could not obtain their source code, so we compare with the numbers they reported in their paper on the same data sets and similar machines.

\begin{figure}[!t]
\vspace{-5pt}
    \centering
    \subfloat[Total running time of DPC algorithms\label{fig:dpc-comparison-a}]{{\includegraphics[width=0.85\columnwidth]{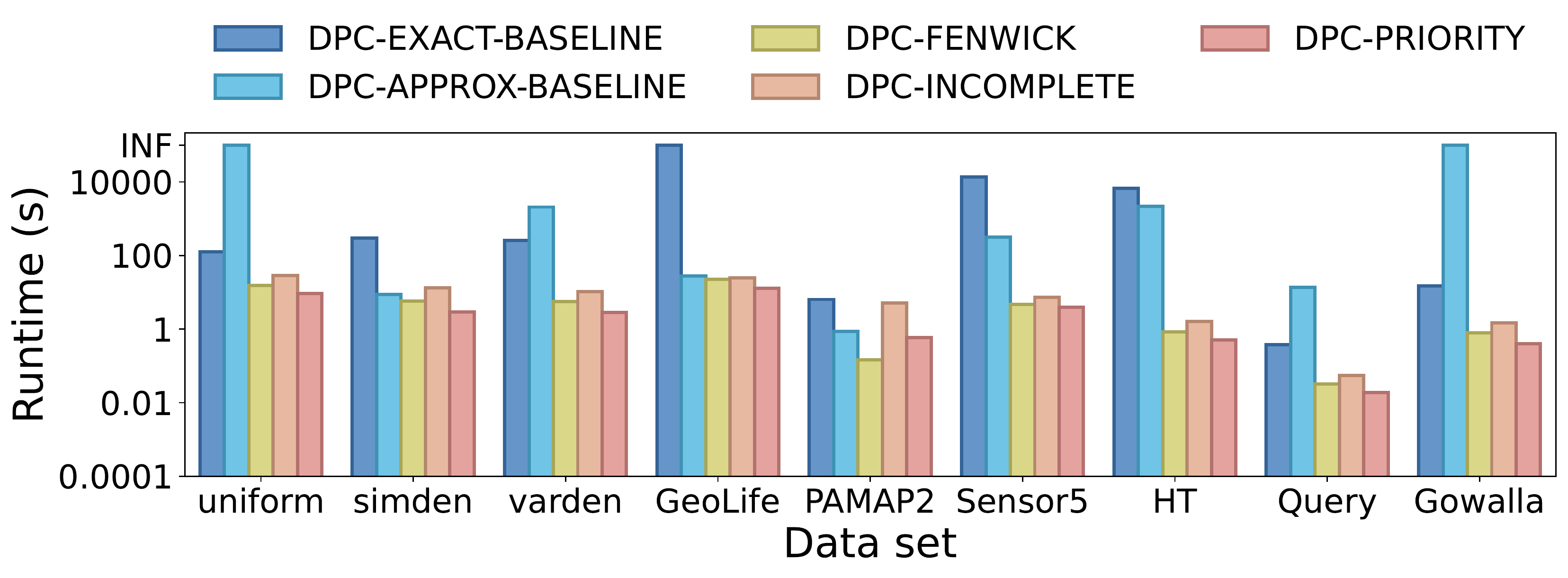}}}
    \quad
    \subfloat[Running time of density computation\label{fig:dpc-comparison-b}]{{\includegraphics[width=0.85\columnwidth]{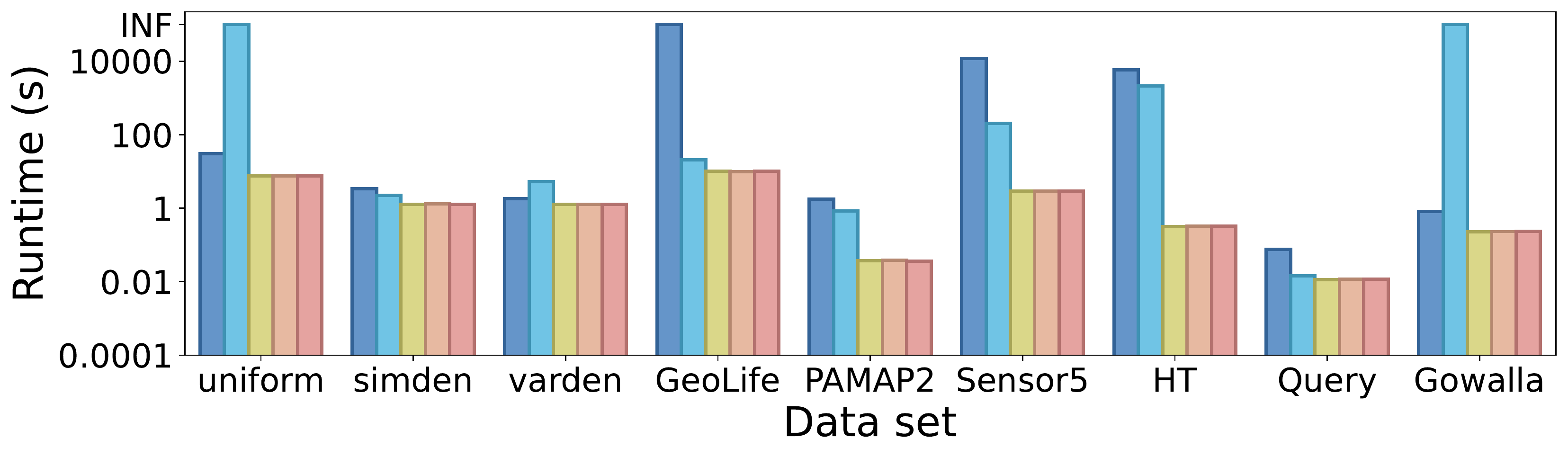}}}
    \quad
    \subfloat[Running time of dependent point finding\label{fig:dpc-comparison-c}]{{\includegraphics[width=0.85\columnwidth]{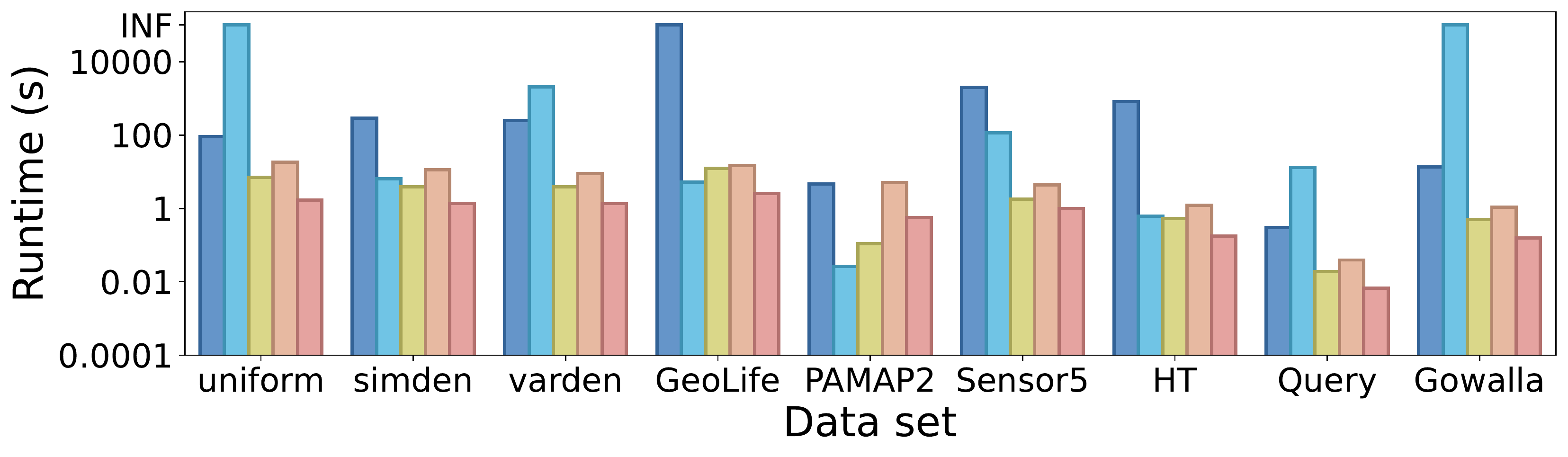}}}
    \caption{Running times (seconds) of DPC algorithms in log-scale. All algorithms are run on a 30-core machine with hyper-threading. Some algorithms time out and do not terminate within 48 hours. These algorithms have time ``INF". Our proposed dependent point finding algorithms and density computation optimizations achieve significant improvement in comparison to other methods. For uniform, simden, and varden, we used $n=10^7$.} \label{fig:dpc-comparison}%
\end{figure}

\begin{table*}[t]
\vspace{-5pt}
 \setlength{\tabcolsep}{3pt}
\footnotesize
\centering
 \begin{tabular}{l |ccc|ccc|ccc|ccc|ccc} 
 \toprule
  \multicolumn{1}{l|}{Algorithm} & \multicolumn{3}{c|}{\algname{dpc-exact-baseline}} & \multicolumn{3}{c|}{\algname{dpc-approx-baseline}} & \multicolumn{3}{c|}{\algname{dpc-Fenwick}} & \multicolumn{3}{c|}{\algname{dpc-incomplete}} & \multicolumn{3}{c}{\algname{dpc-priority}} \\ %
  \midrule
  Datasets & density & dep. & total & density & dep. & total & density & dep. & total & density & dep. & total & density & dep.  & total\\
 \midrule
 \emph{uniform2} & $30.70$ & $91.30$ & 125.44 & -- & -- & -- & $7.65$ & $7.07$ & 15.43 & \textbf{7.58} & $18.26$ & 28.50 & 7.59 & \textbf{1.69}&  \textbf{9.30}  \\
 \emph{simden2} & $3.39$ & $290.30$ & 296.81 & $2.23$ & $6.36$ & 8.74 & $1.29$ & $3.86$ & 5.85 & $1.31$ & $11.27$ & 13.12 & \textbf{1.27} & \textbf{1.39}&  \textbf{2.94} \\
 \emph{varden2} & $1.82$ & $250.23$ & 256.28 & $5.25$ & $2072.96$ & 2078.41 & $1.28$ & $3.87$&  5.63 & \textbf{1.26} & $8.93$ & 10.60 & $1.28$ & \textbf{1.35} & \textbf{2.86} \\
 \emph{GeoLife} & -- & -- & -- & $21.08$ & $5.19$ & 28.12 & $10.20$ & $12.25$ & 22.48 & \textbf{10.04} & $14.95$ &  25.03 & $10.18$ & \textbf{2.59} &  \textbf{12.80}  \\
 \emph{PAMAP2} & $1.76$ & $4.65$ &  6.41 & 0.83 & \textbf{0.026} & 0.86  & \textbf{0.037} & 0.11 & \textbf{0.15} & $0.052$ & $5.13$ & 5.18 & $0.037$ & $0.56$ &  0.59 \\
 \emph{Sensor} & $11850.20$ & $2000.41$ & 13852.72 & $202.95$ & $115.50$ & 318.60 & $2.97$ & $1.77$ & 4.75  & \textbf{2.94} & $4.33$ & 7.29 & $2.95$ & \textbf{0.98} & \textbf{3.94}  \\
 \emph{HT} & $5836.56$ & $814.50$ & 6652.43 & $2144.31$ & $0.61$ & 2144.93 & \textbf{0.31} & $0.52$& 0.85  & $0.46$ & $1.21$ & 1.63 & $0.32$ & \textbf{0.17} &  \textbf{0.51}\\
 \emph{Query} & 0.08 & 0.30 & 0.38 & 0.014 & 13.3 & 13.58 & \textbf{0.01} & 0.019 & 0.03 & 0.01 & 0.039 & 0.05 & 0.01 & \textbf{0.007} & \textbf{0.02}\\
 \emph{Gowalla} & 0.82 & 13.57 & 14.72 & -- & -- & -- & \textbf{0.23} & 0.49 & 0.78 & 0.23 & 1.09 & 1.48 & 0.24 & \textbf{0.16} & \textbf{0.40}\\
 \bottomrule
\end{tabular}
\caption{\label{tab:dpc-comparison} The running times (seconds) of the 5 DPC algorithms on real-world and synthetic data sets, decomposed into the density computation step (density) and the dependent point finding step (dep.). "--" means that the algorithm did not terminate within 48 hours. 
}
\vspace{2pt}
\end{table*}

\subsection{Runtime Comparison}
In this subsection, we compare the DPC algorithms' overall performance, and the runtimes of the density computation task separately from the dependent point finding task in order to study the effectiveness of our proposed optimizations for those tasks. The single linkage clustering task is not studied separately as it takes up a negligible percentage of the overall runtime in all algorithms. 
Figure~\ref{fig:dpc-comparison} 
and Table~\ref{tab:dpc-comparison} 
show the runtime comparison across the DPC algorithms.

As shown in \Cref{tab:dpc-comparison}, the portion of time that density computation and dependent point finding take  is highly dependent on the data set and $d_\text{cut}$ parameters we use. For our experiments, $d_\text{cut}$ is chosen such that the computed density values are nonzero but are much less than $n$.

\myparagraph{Comparison across exact DPC algorithms}
From \Cref{fig:dpc-comparison}, we can see that all of our proposed algorithms consistently outperform \algname{dpc-exact-baseline} on all data sets.
Only \algname{dpc-incomplete} is slightly slower than \algname{dpc-exact-baseline} for PAMAP2 in the dependent point finding step. %
Overall, \algname{dpc-priority} is the fastest on almost all data sets, and achieves a 10.8--13169x speedup over \algname{dpc-exact-baseline} (\Cref{fig:dpc-comparison-a}).

On a single thread, \algname{dpc-priority} also outperforms \citet{Rasool20}'s state-of-the-art sequential exact DPC algorithm. \citet{Rasool20} reported their sequential R-tree based algorithm's running time for Query and Gowalla. To compare with their results, we performed experiments on Query and Gowalla using a machine with the same processor specifications as the one they used, and found that our algorithms are 1.1--1.4x faster.

Our optimized density computation method from \Cref{sec:par-other} (which is the same for all three of our algorithms) outperforms \algname{dpc-exact-baseline}'s density computation step by 1.4--18586.3x, with a geometric mean of 31.5x (\Cref{fig:dpc-comparison-b}).
Our method is faster than \algname{dpc-exact-baseline} because we do not need to iterate over all points in the range. Moreover, we pre-allocate memory for all nodes in our \kdt, while the nodes in \algname{dpc-exact-baseline}'s \kdt are allocated dynamically, which can lead to more cache misses. %

For dependent point finding (\Cref{fig:dpc-comparison-c}), \algname{dpc-Fenwick} outperforms \algname{dpc-exact-baseline} by 12.9--1551.7x, with a geometric mean of 81.9x.  \algname{dpc-incomplete} achieves a speedup of 0.9--675.9x, with a geometric mean of 23.6x.
\algname{dpc-priority} attains a speedup of 8.3--4666.3x, with a geometric mean of 168.7x. 
Our fully parallel algorithms \algname{dpc-Fenwick} and \algname{dpc-priority} are faster than \algname{dpc-exact-baseline} mainly because they can find dependent points for all points in parallel. Our \algname{dpc-incomplete} algorithm, which inserts points iteratively like \algname{dpc-exact-baseline}, is still faster because our \kdt is more balanced, does not need to insert points, and has a more cache-friendly layout.

Among our new algorithms, \algname{dpc-Fenwick} and \algname{dpc-priority} are faster than \algname{dpc-incomplete} because the former two are fully parallel. \algname{dpc-priority} is faster than \algname{dpc-Fenwick} on most data sets, due to its lower average-case work bound, but \algname{dpc-Fenwick} can sometimes be faster (e.g., on PAMAP2) depending on the data set distribution. 

\myparagraph{Comparison with approximate DPC baseline}
\algname{dpc-priority} is able to achieve running times that are superior to \algname{dpc-approx-baseline} on all data sets. \algname{dpc-Fenwick} and \algname{dpc-incomplete} can also achieve competitive results when compared to \algname{dpc-approx-baseline}. Across all data sets, our optimized density computation method attains a 1.7--6828.5x speedup over \algname{dpc-approx-baseline}; the geometric mean speedup is 17.6x (\Cref{fig:dpc-comparison-b}). 

Considering just the dependent point finding step (\Cref{fig:dpc-comparison-c}),
\algname{dpc-Fenwick} outperforms \algname{dpc-approx-baseline} by 0.2--536.2x, with a geometric mean of 3.4x; \algname{dpc-incomplete} outperforms \algname{dpc-approx-baseline}  by 0.005--232.2x, with a geometric mean of 1.0x; and \algname{dpc-priority} outperforms \algname{dpc-approx-baseline} by 0.05--1534.1x, with a geometric mean of 6.7x. The range of speedups varies significantly across data sets primarily because \algname{dpc-approx-baseline}'s performance is highly dependent on the distribution of points on each data set. \algname{dpc-priority}'s dependent point finding step is only slower than that of \algname{dpc-approx-baseline} on one data set (PAMAP2), and achieves considerable speedup on all others. 

In total (\Cref{fig:dpc-comparison-a}), \algname{dpc-priority} achieves a 1.5--4206x speedup against \algname{dpc-approx-baseline}, with a geometric mean speedup of 55.4x. 
\algname{dpc-Fenwick} achieves a 1.3-2523x speedup over \algname{dpc-approx-baseline}, with a geometric mean speedup of 43.7x. 
\algname{dpc-incomplete} obtains a 0.7--1316x speedup against \algname{dpc-approx-baseline}, with a geometric mean speedup of 16.8x.

\myparagraph{Effect of parameters on running time}
 $\delta_\text{min}$ is only used in Step~\ref{item:single-linkage} of DPC. Since Step~\ref{item:single-linkage} takes negligible percentage of the overall runtime, using different $\delta_\text{min}$ values has little effect on the total runtime.
Increasing $\rho_\text{min}$ increases the number of noise points, and hence the overall running time decreases because noise points are skipped in Steps~\ref{item:dep-point} and~\ref{item:single-linkage}.
 For $d_\text{cut}$, the effect on total running time depends on the distribution of data sets. We show the effect of $d_\text{cut}$ in 
\iffull
\Cref{sec:dcut}. 
\fi
In general, a higher $d_\text{cut}$ leads to increased running time. 

\begin{figure}[!t]
\vspace{-5pt}
    \centering
    \subfloat{{\includegraphics[width=0.8\linewidth]{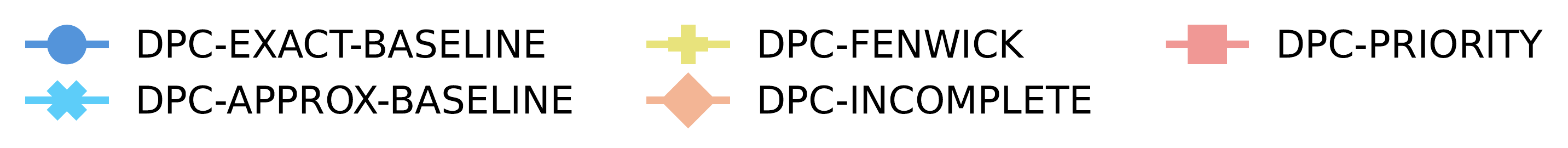}}}
    \setcounter{subfigure}{0}
    \quad
    \subfloat[Running time (seconds) across data sets of different sizes.\label{fig:dpc-scalability-n}]{{\includegraphics[width=0.44\linewidth]{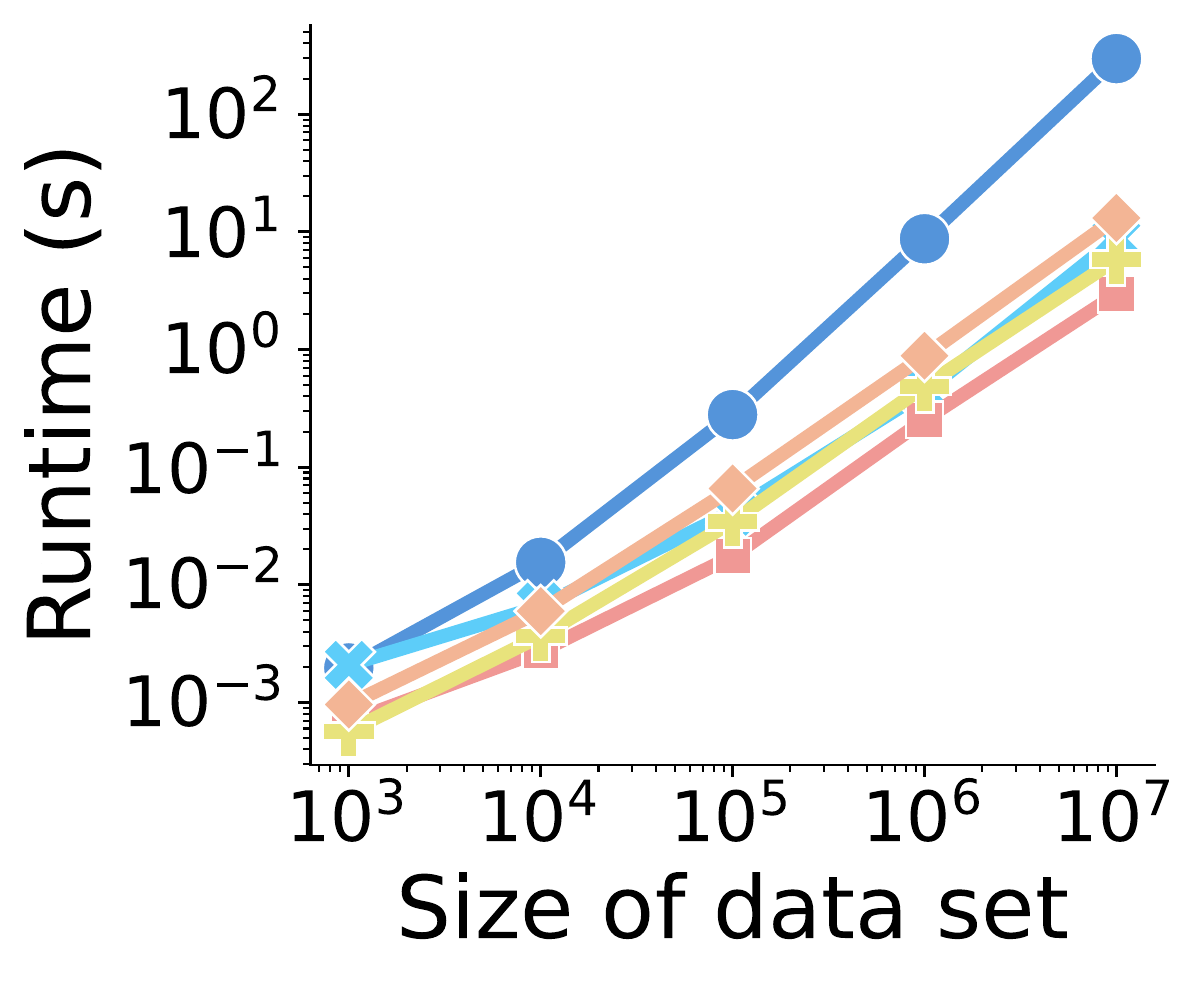}}}
    \quad
    \subfloat[Speedup across different numbers of threads.\label{fig:dpc-scalability-thread}]{{\includegraphics[width=0.44\linewidth]{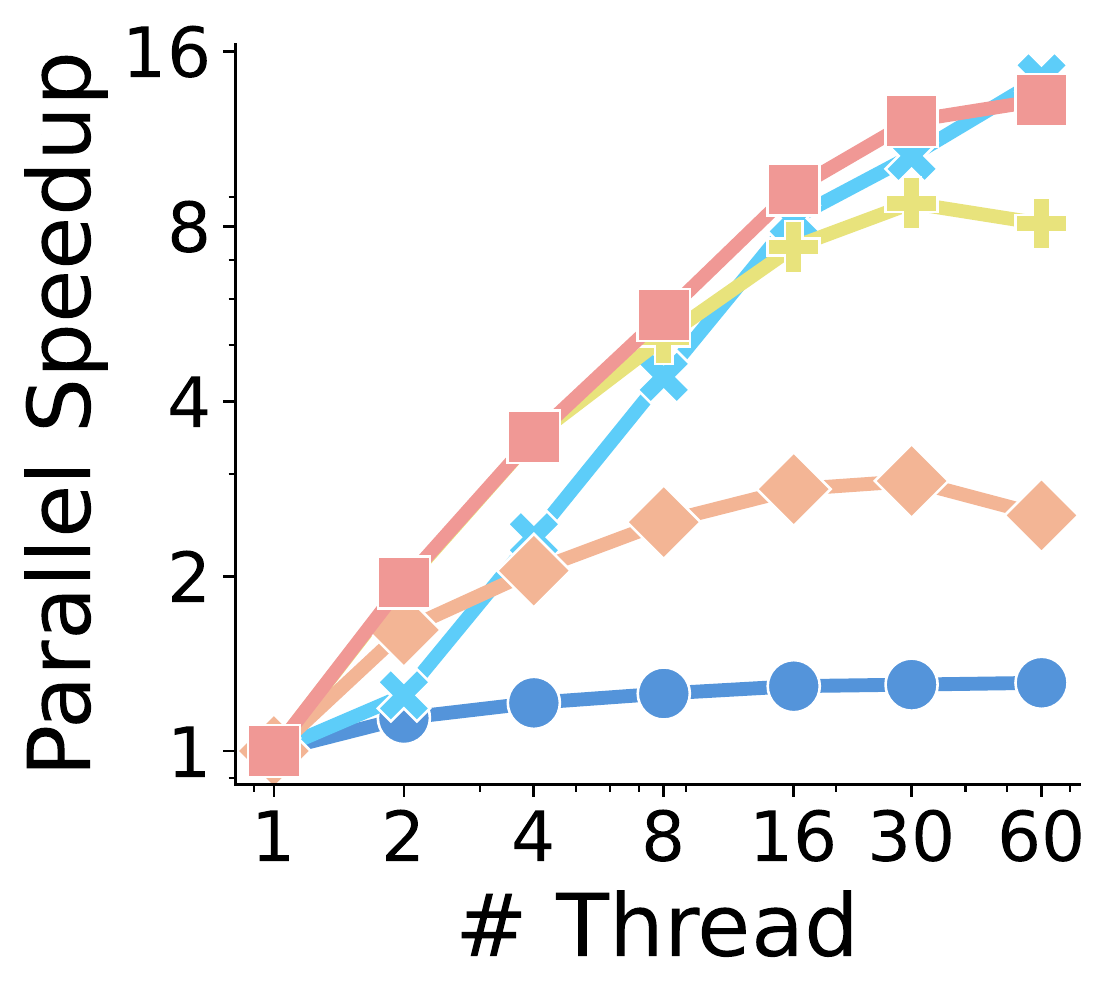}}}
    \caption{Running time of DPC algorithms on \emph{simden} data sets of different sizes and the parallel speedup of the DPC algorithms across different number of threads (``60 threads" means 30 cores with two-way hyper-threading) on the \emph{simden} data set of size $10^7$. All axes use logarithmic  scale.}\label{fig:dpc-scalability} %
\end{figure}

\subsection{Scalability Analysis}

We analyze the scalability of our algorithms by performing experiments on synthetic datasets of varying sizes and running the algorithms on different numbers of threads. We use datasets generated by \emph{simden} for scalability analysis because \algname{dpc-approx-baseline}, when running on a single thread, does not terminate for the largest \emph{uniform} and \emph{varden} datasets within 48 hours.

\myparagraph{Scalability over the size of the dataset}
Figure~\ref{fig:dpc-scalability-n} shows the runtime of all DPC algorithms over \emph{simden} datasets of different sizes (from $10^3$ points to $10^7$ points). Our \algname{dpc-priority} outperforms both \algname{dpc-exact-baseline} and \algname{dpc-approx-baseline} for \emph{simden} data sets of all sizes tested. Furthermore, we see that the running time of our proposed algorithms increases much more slowly than \algname{dpc-exact-baseline} as the data size increases. 
We use a linear fit on the logarithm of runntime and $ \log n$ to get the slopes of the lines in \Cref{fig:dpc-scalability-n}.
The slope for  \algname{dpc-exact-baseline} is 1.31, for \algname{dpc-approx-baseline} is 0.94, for \algname{dpc-Fenwick} is 1.02, for \algname{dpc-incomplete} is 1.05, and for \algname{dpc-priority} is 0.94. 
 This demonstrates that our algorithm has superior scalability across different graph sizes.

\myparagraph{Parallel scalability}
Finally, we investigate the parallel scalability of our algorithms. Figure \ref{fig:dpc-scalability-thread} shows that all of our proposed DPC algorithms obtain better parallel scallability than \algname{dpc-exact-baseline}.
\algname{dpc-Fenwick} is able to achieve a 8.8x self-relative speedup and \algname{dpc-priority} achieves a 13.2x self-relative speedup. Both are superior to the 1.3x self-relative speedup attained by \algname{dpc-exact-baseline} and are competitive against the 14.4x self-relative speedup achieved by \algname{dpc-approx-baseline}. \algname{dpc-Fenwick} and \algname{dpc-incomplete} have smaller speedups on 60 hyper-threads than on 30 threads due to the extra overhead of hyper-threading.

\section{Conclusion}
In this paper, we developed efficient parallel algorithms for density peaks clustering, and proved strong work and span bounds for them.
We performed experiments of our algorithms, showing that they outperform previous state-of-the-art DPC algorithms and achieve good parallel scalability on large data sets.
For future work, we are interested in exploring distributed versions of DPC.

\myparagraph{Acknowledgements}
This research is supported by MIT PRIMES, Siebel Scholars program,
DOE Early Career Award \#DE-SC0018947,
NSF CAREER Award \#CCF-1845763, Google Faculty Research Award, Google Research Scholar Award, cloud computing credits from Google-MIT,
FinTech@CSAIL Initiative, DARPA
SDH Award \#HR0011-18-3-0007, and Applications Driving Architectures
(ADA) Research Center, a JUMP Center co-sponsored by SRC and DARPA.

\clearpage
\bibliographystyle{plainnat}
\bibliography{references}

\iffull
\appendix

\section{Range Query Complexity on Priority Search \kdt{s}}\label{sec:priority-range-query}
A priority search \kdt can be used to efficiently answer a priority range query, which is defined as follows.
Note that we do not use the priority range search for our DPC algorithms, but we believe it can potentially be useful in other applications.

\begin{definition}
    Given a query range $R_q\subseteq \mathbb{R}^d$, a priority threshold $\gamma_q$, and a point set $P$, a \textbf{priority range search} asks for all points $x_i$ in $P$ such that $x_i\in R_q$ and $\gamma_i > \gamma_q$.  
\end{definition}

A priority range query can be solved using a standard \kdt by querying the set of points inside $R_q$ and filtering for points that satisfies the priority constraint. 

This can be optimized by using a priority search \kdt. When performing a priority range query on a priority search \kdt $T$, we only visit nodes whose cell intersects the query region $R_q$ and has a $\gamma$  value greater than the cutoff value $\gamma_q$. Let $T_q$ represent the subset of nodes with $\gamma$ value higher than $\gamma_q$. $T_q$ is necessarily a connected subtree that forms an upper portion of $T$, as illustrated by Figure \ref{fig:priority-kdt} and proved in \Cref{sec:priority-kd-tree}. Therefore, performing a priority range search on $T$ is equivalent to performing a normal range search on $T_q$. 

\myparagraph{Analysis}
If $R_q$ is an \textit{axis-parallel hyper-rectangular region}%
, then a meaningful complexity bound can be established for priority range query on a priority search \kdt $T$. All cells visited during the query operation must be in $T_q$. There are two types of such cells:

\begin{enumerate}
    \item A cell that intersects $R_q$ but is not completely inside $R_q$: 
    It is well known that the number of such cells in $T$ is bounded by $\BigO{n^{1-\frac{1}{d}}}$ \cite{CLRS}.
    \item A cell that is completely inside $R_q$: 
    Each such cell must contain a unique point $x_i$ satisfying $x_i\in R_q$ and $\gamma_i > \gamma_q$. Hence the number of cells of this type is bounded by the number of points inside $R_q$ with a priority value $>\gamma_q$. Let $Q$ represent the set of such points.\label{item:type2-cell}%
\end{enumerate}

In sum, the total number of cells visited by a priority range query operation is bounded by $\BigO{n^{1-\frac{1}{d}}+|Q|}$. 

Note that this proof cannot be applied to a max \kdt because each cell in a max \kdt is not uniquely associated with a point, as assumed in the proof in order to bound the number of Type \ref{item:type2-cell} cells traversed.

\section{$K$-Nearest Neighbor Query Complexity on Priority Search \kdt{s}}\label{sec:complexity}
In this section, we provide an analysis of the average-case complexity of a $K$-nearest neighbor query  on an incomplete \kdt. We first note that a priority $K$-nearest neighbor query on a priority search \kdt can be reduced to a normal $K$-nearest neighbor query on an incomplete \kdt, as discussed in Section~\ref{sec:priority-search-kdt-knn}. 

As a reminder, the priority $K$-nearest neighbor query is formally defined as follows. 

\begin{definition}
    Given a query point $x_q\in \mathbb{R}^d$, a point set $P$, and a distance measure $D$, the set of priority $K$-nearest neighbors is the set of $K$ points $\{x_1, x_2, \ldots, x_i, \ldots, x_k\}$ such that $\lambda_i > \lambda_q$ for all $i\in [1, K]$ and $x_i$ is the $i^\text{th}$ closest point to $x_q$ in this set as measured by $D$. 
\end{definition}

We define a general incomplete \kdt to be a \kdt constructed from $P=\{x_1, x_2, \ldots, x_n\}$ in $d$-dimensional space such that $x_i$ is only active if $x_i\in P_q$, where $P_q\subseteq P$. Thus, performing a priority $K$-nearest neighbor query for query point $x_q$ with priority value $\lambda_q$ is equivalent to performing a $K$-nearest neighbor query on an incomplete \kdt with the active point set $P_q = \{x_i \mid \lambda_i>\lambda_q\text{ and } x_i\in P\}$.
We define an \defn{active node} to be a node with at least one active point in its cell.

We assume the incomplete \kdt splits along its cell's widest dimension at each level. We further assume that the incomplete \kdt stores all of its points in its leaf nodes. The incomplete \kdt can also store points at its internal nodes, as is the case for the priority search \kdt. Although assuming all points to be stored at the leaf nodes simplifies the proof, this storage method does not affect the validity of the proof. The reason is that a \kdt storing points in its internal nodes can be converted into a \kdt that only stores point in the leaf nodes by pushing the stored points down the branches of the \kdt until they are all at leaf nodes. Any traversal through the new \kdt mirrors a traversal through the original \kdt, except that traversal through the original \kdt can terminate in fewer steps---since points are stored higher up on the \kdt, we do not always need to traverse to the leaves of the \kdt to find the point that we are looking for. Hence, storing points in internal nodes as well as leaf nodes does not make the \kdt's performance worse.

Although we prove the general complexity result for $K$-nearest neighbor queries, our DPC algorithm only makes use of the nearest neighbor query, which is the special case when $K=1$. Also note that the $k$ in \kdt is different from the $K$ in $K$-nearest neighbor. 

The algorithm for computing the $K$-nearest neighbors is similar to that of computing the nearest neighbor as described in Section \ref{sec:relevant-technique}.
To find the $K$-nearest neighbors of a point $x$, we first traverse down the incomplete \kdt to find the leaf node that contains the point $x$. Then, in the backtracking process, we search the sibling subtrees. Let $x$'s distance to the current $K^\text{th}$ nearest neighbor candidate of $x$ be represented by $L_K$. We prune the search of any subtree whose cell is farther than $L_K$ away from $x$. We also prune the search of a subtree if it does not contain an active point.

Our analysis follows in a similar spirit to \citet{Friedman77}'s proof of the  $\BigO{K\log n}$ average-case complexity for $K$-nearest neighbor query on a normal \kdt. We show that finding the $K$-nearest neighbors of some query point $x_q$ on an incomplete \kdt takes $\BigO{K\log n}$ average-case work.

We separate the analysis based on the size of $P_q$.

\myparagraph{Case 1: $|P_q|=\BigO{K\log n}$}

In this case, we can simply find the $K$-nearest neighbors of a point $x_q$ by searching through all active points in the incomplete \kdt in $\BigO{K\log n}$ work. 

\myparagraph{Case 2: $|P_q|=\omega{(K\log n)}$}

To analyze this case, we make some similar assumptions as \citet{Friedman77}. 
We list our notation in \Cref{tbl:proofnotation}.
Let $\mu$ be the probability density function that all $x_i\in P$ are sampled from. Similarly, let $\mu_q$ be the probability density function that all $x_i\in P_q$ are sampled from. We use $\mu(x)$ and $\mu_q(x)$ to denote the probability density of functions $\mu$ and $\mu_q$, respectively, at a generic point $x$. 

Let $\epsilon_1$ and $\epsilon_2$ be some small constants greater than $0$.
Our assumptions are as follows: 
\begin{enumerate}  
    \item The dimension $d$ is a constant. 
    \item The incomplete \kdt has \defn{compact cells}. Specifically, cells with $\le (1+\epsilon_2)K$ active points are near hyper-cubical in shape, where the ratio of the longest side to the shortest side is bounded by $(1+\epsilon_1)$. 
    \item $|P_q|$ is sufficiently large such that $\mu_q$ can be considered locally uniform. Specifically, $\mu_q$ is assumed to be constant in expectation over all samples of $P_q$ within a hyper-spherical region centered at the query point with radius $\le 3\sqrt{d}\cdot \text{diag}(R^K_q)$, where $\text{diag}(R^K_q)$ is the diagonal length of the cell $R^K_q$ corresponding to the smallest subtree that contains $x_q$ and $\ge K$ active points. For query point $x_q$, we represent this constant by $\mu_q(x_q)$. 
    \item We further constrain the local uniformity of $\mu_q$ by requiring that for each node with $\mathcal{K}<K$ active points, its sibling node has at most $(1+\epsilon_2)\mathcal{K}$ points in expectation.
\end{enumerate}

We will first bound the volume of the region that we need to traverse in the backtracking stage, and then bound the number of cells that intersect with this region.

\begin{table}[!t]
\footnotesize
\centering
 \begin{tabular}{l c } 
 \toprule
Symbol & Description\\ %
 \midrule
 $x_q$ & query point \\
 $\mu$ & probability density function of $P$ \\
$\mu_q$ & probability density function of $P_q$ \\
$N_q$ & leaf node of $x_q$ \\
$R_q$ & cell of $N_q$ \\
$N_q^{K}$ & the smallest subtree that contains $N_q$ \\
 & and contains $\ge K$ active points \\
$R_q^{K}$ & cell of $N_q^{K}$ \\
$[ \cdot ]$ & expected value over all sampling of points \\
$\mu_q(x_q)$ & the local constant density in  $R_q^{K}$\\
$V(\cdot)$ & the volume of a region \\ 
$u(\cdot)$ & the probability mass of a region \\
$\text{diag}(\cdot)$ & diagonal length\\
$U_q^K$ & the smallest hyper-sphere that encloses $R_q^K$ \\
$S_q^{K}$ & a hyper-cube concentric with $S_q^K$ with side length $2\cdot\text{diag}(R_q^{K})$\\
$B_q^{K}$ & a hyper-sphere concentric with $S_q^{K}$\\
& with radius $3\sqrt{d}\cdot \text{diag}(R_q^{K})$\\
 \bottomrule
\end{tabular}
\caption{Notation used in the proofs.}\label{tbl:proofnotation}
\end{table}

Let $N_q$ represent a leaf node with cell $R_q$ such that our query point $x_q$ is contained within $R_q$. We define $N^K_q$ to be the smallest subtree that contains $N_q$ and contains at least $K$ active points; let the cell of $N^K_q$ be represented by $R^K_q$. The children cells of $R^K_q$ are adjacent compact cells. Since the child subtree containing $x_q$ has fewer than $K$ active points, the sibling cell contains $\le K(1+\epsilon_2)$ active points in expectation based on our assumptions.
Thus, the expected number of active points in $N^K_q$ is $[\text{size}(N^K_q)] \le (2+\epsilon_2)K$, where the square bracket $[ \cdot ]$ indicates taking the expected value over all sampling of points. 

We define $V(R_q^K)$ to be the volume of $R^K_q$, and define its probability mass to be
\begin{align*}
    u(R_q^K) = \int_{x\in R_q^K} \mu_q(x) dx. 
\end{align*}

Let $U_q^k$ be the smallest hyper-sphere entered at $q$ that encloses $R_q^K$. 
Due to the bounded aspect ratio assumption, the volume of $U_q^K$ is at most a constant times larger than that of $R_q^K$. We denote this constant by $C(d,\epsilon_1)$. 
Based on the local uniformity assumption, the number of active points in $U_q^K$ is a  factor of $C(d,\epsilon_1)$ larger than in $R_q^K$ in expectation, so $[\text{size}(U^K_q)] \le (2+\epsilon_2)C(d,\epsilon_1)K$.

The probability mass of  $U_q^K$ is similarly defined to be 
\begin{align*}
u(U_q^K)= \int_{x\in U_q^K} \mu_q(x) dx
\end{align*}

This follows a beta distribution~\cite{Fukunaga73, fraser1956nonparametric} with mean 

\begin{align*}
[u(U_q^K)] = \frac{[\text{size}(U_q^K)]}{|P_q|+1} \leq \frac{(2+\epsilon_2)C(d,\epsilon_1)K}{|P_q|+1}.
\end{align*}

As $R_q^K$ is completely contained in $U_q^K$, we have
\begin{align*}
[u(R_q^K)] \leq [u(U_q^K)] \leq \frac{(2+\epsilon_2)C(d,\epsilon_1)K}{|P_q|+1}.
\end{align*}

Note that $u(R_q^K)$ can be expressed in another way: 
\begin{align*}
    u(R_q^K) &= \int_{x\in R^K_q} \mu_q(x) dx \\
    &= \mu_q(x_q) V(R_q^K).
\end{align*}
because $\mu_q$ is assumed to be locally uniform. We combine the two relations to get
\begin{align*}
    \mu_q(x_q) [V(R_q^K)] &\le \frac{(2+\epsilon_2)C(d,\epsilon_1)K}{|P_q|+1} \\
    [V(R_q^K)] &\le \frac{(2+\epsilon_2)C(d,\epsilon_1)K}{(|P_q|+1)\mu_q(x_q)}.
\end{align*}

\begin{figure}
    \centering
    \includegraphics[width=\linewidth]{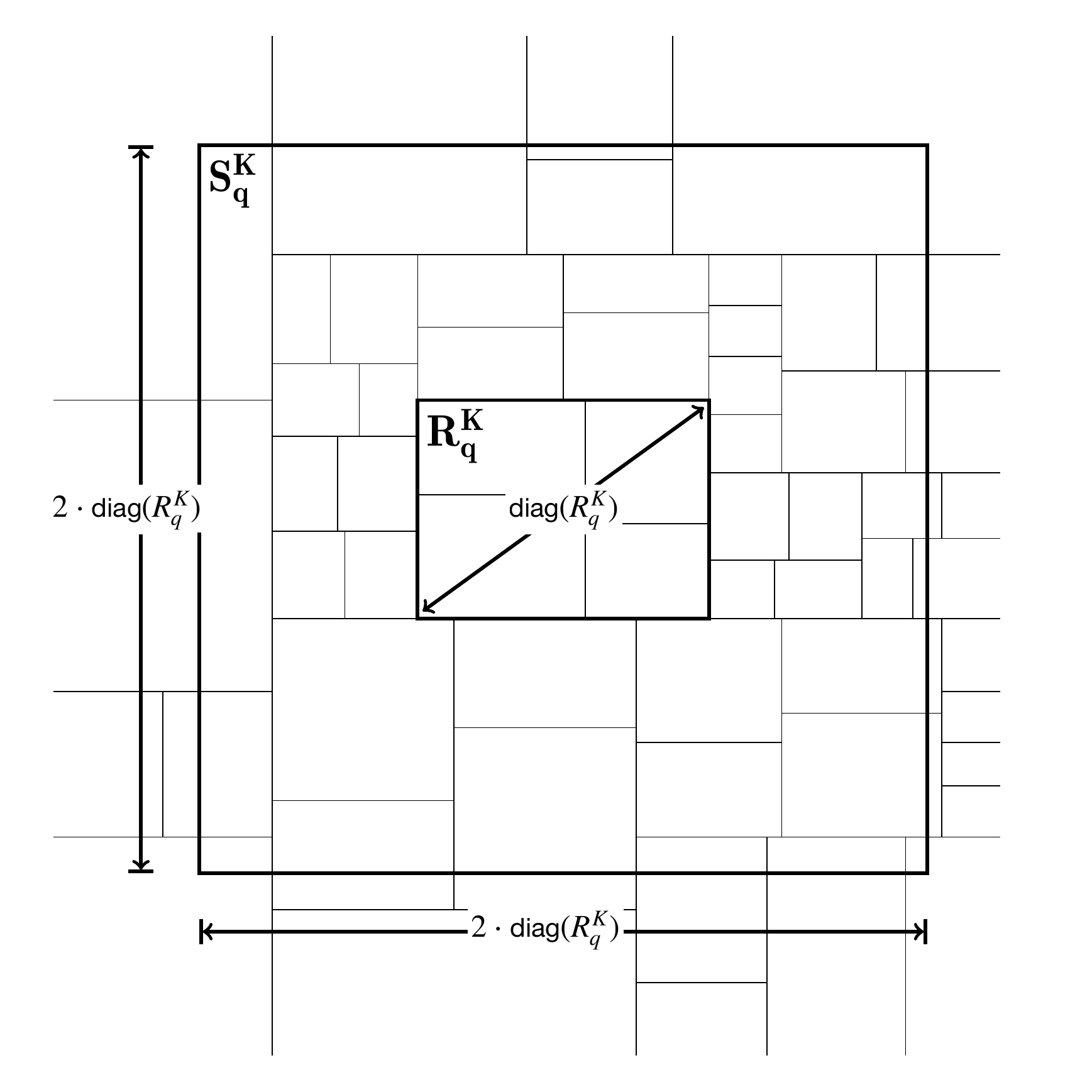}
    \caption{Example division of space by an incomplete \kdt. Each rectangular region with unbolded sides represent a leaf cell. } \label{fig:appendix}
\end{figure}

Consider the \kdt query procedure for finding the $K$-nearest neighbors of $x_q$. In the first step, we traverse down the incomplete \kdt to find $N_q$, the leaf node containing $x_q$. Then, we backtrack up the incomplete \kdt, visiting neighboring sibling subtrees that contain active points. 
Therefore, once we backtrack to node $N_q^K$, we are guaranteed to have searched through at least $K$ active points. The maximum possible distance between $x_q$ and its $K^\text{th}$ nearest neighbor is bounded by the diagonal of the hyper-rectangle $R^K_q$. Let this diagonal length be denoted by $\text{diag}(R^K_q)$ and let $S^K_q$ represent a hyper-cube with side length $2\text{diag}(R^K_q)$, centered at $x_q$. See \Cref{fig:appendix} for an illustration.
When we continue to traverse up the incomplete \kdt, we only need to search through cells that intersect $S^K_q$, since cells outside of $S^K_q$ are necessarily $> \text{diag}(R^K_q)$ away from $x_q$. 

We want to bound the number of active leaf cells in the incomplete \kdt that intersect with $S_q^K$ (see \Cref{fig:appendix}). Note that each of these leaf cells are near hyper-cubical in shape. However, these leaf cells can have different side lengths because $\mu$ (which determines the sizes of leaf cells) is not guaranteed to be constant. 

There are two types of leaf cells. We use $a_\text{long}$ to denote a cell's longest side length and $a_\text{short}$ to denote its shortest side length. By our assumption, $a_\text{long} \le a_\text{short}(1+\epsilon_1)$. 

\myparagraph{Type 1: the leaf cell's $a_\text{long}>2\cdot\text{diag}(R^K_q)$}

Because $a_\text{long}>2\cdot\text{diag}(R^K_q)$, we have $a_\text{short}>\frac{2\cdot\text{diag}(R^K_q)}{1+\epsilon_1}$. Since the longest side of the leaf cell is longer than the side length of $S^K_q$, it must intersect at least one of $S^K_q$'s $2d$ hyper-surfaces. Leaf cells must be non-overlapping. Hence the number of Type 1 leaf cells that intersect a particular hyper-surface of $S^K_q$ is bounded by
\begin{align*}
    &\left(\frac{\text{side length of $S^K_q$}}{\text{a leaf cell's shortest possible side length}}\right)^d \\
    =& \left(\frac{2\cdot\text{diag}(R^K_q)}{a_\text{short}}\right)^d < \left(1+\epsilon_1\right)^d.
\end{align*}

The total number of Type 1 leaf cells intersecting $S^K_q$ is, therefore, bounded by $2d \left(1+\epsilon_1\right)^d$, which is a constant.

\myparagraph{Type 2: the leaf cell's $a_\text{long}\le 2\cdot\text{diag}(R^K_q)$}

If we let $B_q^K$ be the hyper-sphere concentric with $S^K_q$ with a radius of $3\sqrt{d}\cdot \text{diag}(R^K_q)$, then all Type 2 leaf cells intersecting $S_q^K$ must be contained within $B_q^K$. The radius of $B_q^K$ is proportional to the diagonal of $R^K_q$. In addition, $R^K_q$ has a bounded aspect ratio (the ratio between its longest side and shortest side) because its children cells are compact cells. Therefore, we know that the expected volume of $B_q^K$ is bounded by a constant times the expected volume of $R^K_q$. 
Specifically,
\begin{align*}
[V(B^K_q)] &\le G(d,\epsilon_1) [V(R^K_q)] \\
&\le \frac{(2+\epsilon_2) C(d,\epsilon_1)G(d,\epsilon_1)K}{(|P_q|+1)\mu_q(x_q)}\\
& = \frac{\mathcal{C}K}{(|P_q|+1)\mu_q(x_q)}\\
\end{align*}
where $G(d,\epsilon_1)$ is a constant dependent only on the number of dimensions~\cite{Friedman77} and $\epsilon_1$ and  $\mathcal{C} = (2+\epsilon_2)C(d,\epsilon_1) G(d,\epsilon_1)$. 

Now we bound the number of active leaf cells contained inside $B_q^K$. Each active leaf cell contains an active point. Hence, the number of active leaf cells contained in $B_q^K$ is bounded by the number of active points inside $B_q^K$,
which we denote by $|B_q^K \cap P_q|$. 

Following the same argument made for $R^K_q$, the probability mass of $B_q^K$ follows a beta distribution. Thus,
\begin{align*}
    u(B^K_q) &= \int_{x\in B_q^K} \mu_q(x) dx \\
    [u(B^K_q)] &= \frac{[|B_q^K \cap P_q|]}{|P_q|+1}. 
\end{align*}

Note that $\mu_q$ can be considered constant within $S_q^K$ and $B_q^K$ based on our assumptions. Hence, the integral is also equal to $\mu_q(x_q) V(B_q^K)$ and we obtain
\begin{align*}
    \mu_q(x_q) [V(B_q^K)] &= \frac{[|B_q^K \cap P_q|]}{|P_q|+1} \\
    [|B_q^K \cap P_q|] &= \mu_q(x_q) (|P_q| + 1)[V(B_q^K)] \\
    &\le \mu_q(x_q) (|P_q|+1) \frac{\mathcal{C}K}{(|P_q|+1) \mu_q(x_q)} \\
    &= \mathcal{C} K.
\end{align*}

Because only the active leaf cells are visited (our search algorithm never visits a subtree that does not contain an active point), the $K$-nearest neighbor query algorithm traverses $\BigO{K}$ expected number of leaf cells. Visiting each leaf cell takes $\BigO{\log n}$ work, thus giving a total average-case query work complexity of $\BigO{K\log n}$.

The $K$-nearest neighbor query can also be performed in parallel, in which case we first traverse down the \kdt to find the leaf node $N_q$ containing the query point $x_q$, and then backtrack up the \kdt to find the smallest subtree containing at least $K$ active points, $N^K_q$. After this, we only need to inspect \kdt cells that intersect the hyper-cubical region $S^K_q$. These cells can be inspected in parallel. As a result, the span complexity of a parallel $K$-nearest neighbor query can be bounded by the height of the incomplete \kdt, which is $\BigO{\log n}$ in the worst case. The average-case work complexity remains $\BigO{K\log n}$  and the worst-case work complexity is $\BigO{n}$.

This analysis bounds both the $K$-nearest neighbor query complexity for the incomplete \kdt used in Section \ref{sec:incomplete-kd-tree} and for the priority search \kdt described in Section \ref{sec:priority-search-kdt}, since performing a $K$-nearest neighbor query on a priority search \kdt is equivalent to performing nearest neighbor query on a normal incomplete \kdt; the equivalence has been discussed in Section \ref{sec:priority-search-kdt-knn}.

\section{Average-Case Density Computation Complexity}\label{sec:range-appendix}

In this section, we prove that our DPC algorithm's density computation takes $\BigO{\min({n(n^{1-\frac{1}{d}}+\varrho_\text{avg})}, n\varrho_\text{avg}\log n)}$ average-case work. $\varrho_\text{avg}$ is the average cubical density, where the cubical density value of each point equals the number of points around it within a hyper-cubical region with a side length of $2d_\text{cut}$ (notice that $\varrho$ is slightly larger than the ordinary density value $\rho$ for every point). The proof follows in the same vein as the proof for the $K$-nearest neighbor query algorithm and use the same notations. We again make the assumption that all points are stored at leaf nodes of our \kdt.
We also assume the following. 
\begin{enumerate}
    \item Leaf cells have a bounded aspect ratio: the ratio of their longest side to their shortest side is bounded by $(1+\epsilon_1)$.
    \item $d_\text{cut}$ is small enough such that the density within a spherical region with radius $3\sqrt{d}\cdot d_\text{cut}$ has uniform density value. This assumption is reasonable because the $d_\text{cut}$ parameter is supposed to measure the local density of each point.
\end{enumerate}

The crux of the proof is then to bound the number of leaf cells intersecting our query region, which in this case is a hyper-cube $S$ with side length $2d_\text{cut}$. We solved this problem while proving the incomplete \kdt's $K$-nearest neighbor query complexity. We follow the same approach. 

Note that the number of leaf cells with longest side lengths larger than $2d_\text{cut}$ is bounded by $\BigO{1}$ as shown in Section~\ref{sec:complexity}. 
Let $B$ be a hyper-sphere concentric with $S$ and has radius $3\sqrt{d}\cdot d_\text{cut}$. All leaf cells intersecting $S$ and have a longest side length less than or equal to $2d_\text{cut}$ must be contained within $B$. The number of leaf cells contained inside $B$ is bounded by the number of points contained inside $B$. Because of the assumption that the density of points within $B$ is constant, we can bound this quantity by a constant $G(d) \varrho$, where $\varrho$ is the number of points inside the original hyper-cube query region with side length $2d_\text{cut}$ and $G(d)$ is a constant dependent on the dimensionality $d$.

Therefore, each density query searches through $\BigO{\varrho}$ leaf cells. Traversing down to each leaf cell needs $\BigO{\log n}$ work because the height of the \kdt is $\BigO{\log n}$. Hence, performing one density query takes $\BigO{\varrho \log n}$ work. Performing all density queries takes 
\begin{align*}
    \BigO{\sum_{x_i\in P} \varrho(x_i) \log n } = \BigO{n \varrho_\text{avg} \log n}
\end{align*} 
work.

Since the worst-case bound is $\BigO{n(n^{1-\frac{1}{d}}+\varrho_\text{avg})}$, the final average-case bound is $\BigO{\min({n(n^{1-\frac{1}{d}}+\varrho_\text{avg})}, n\varrho_\text{avg}\log n)}$. Due to the local uniformity assumption, we can rewrite $\varrho = H(d)\rho$ where $H(d)$ is a dimensionality-dependent constant. So, the average-case bound can also be written as
\begin{align*}
    \BigO{\min({n(n^{1-\frac{1}{d}}+\rho_\text{avg})}, n\rho_\text{avg}\log n)}.
\end{align*}

\begin{figure*}[!th]
    \centering
    \subfloat[Total running time (seconds) of \algname{dpc-priority} as the average percent of points within a radius of $d_\text{cut}$ increases.\label{fig:dpc-dcut-total}]{{\includegraphics[width=0.64\columnwidth]{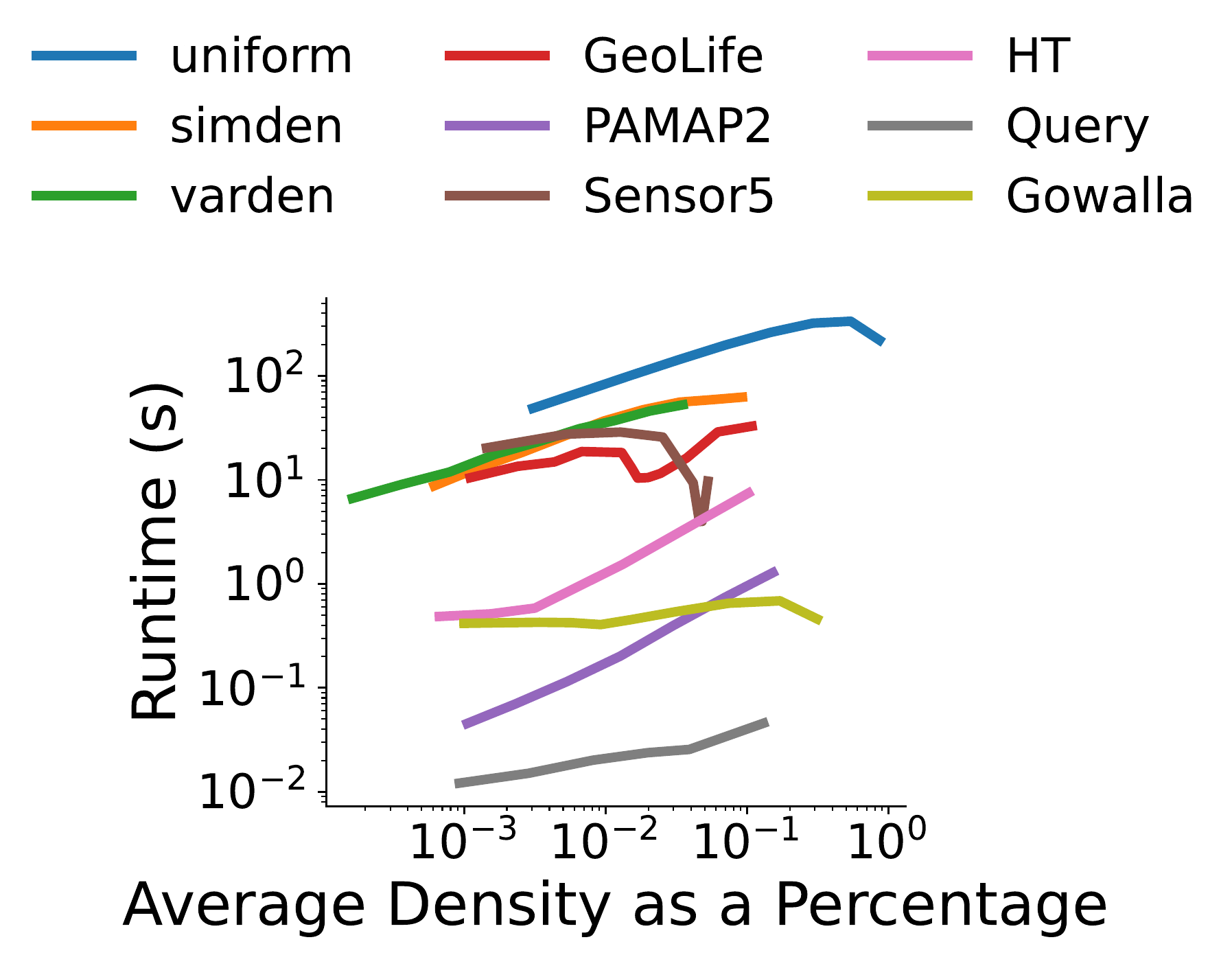}}}
    \quad
    \subfloat[Running time (seconds) of density computation.\label{fig:dpc-dcut-density}]{{\includegraphics[width=0.64\columnwidth]{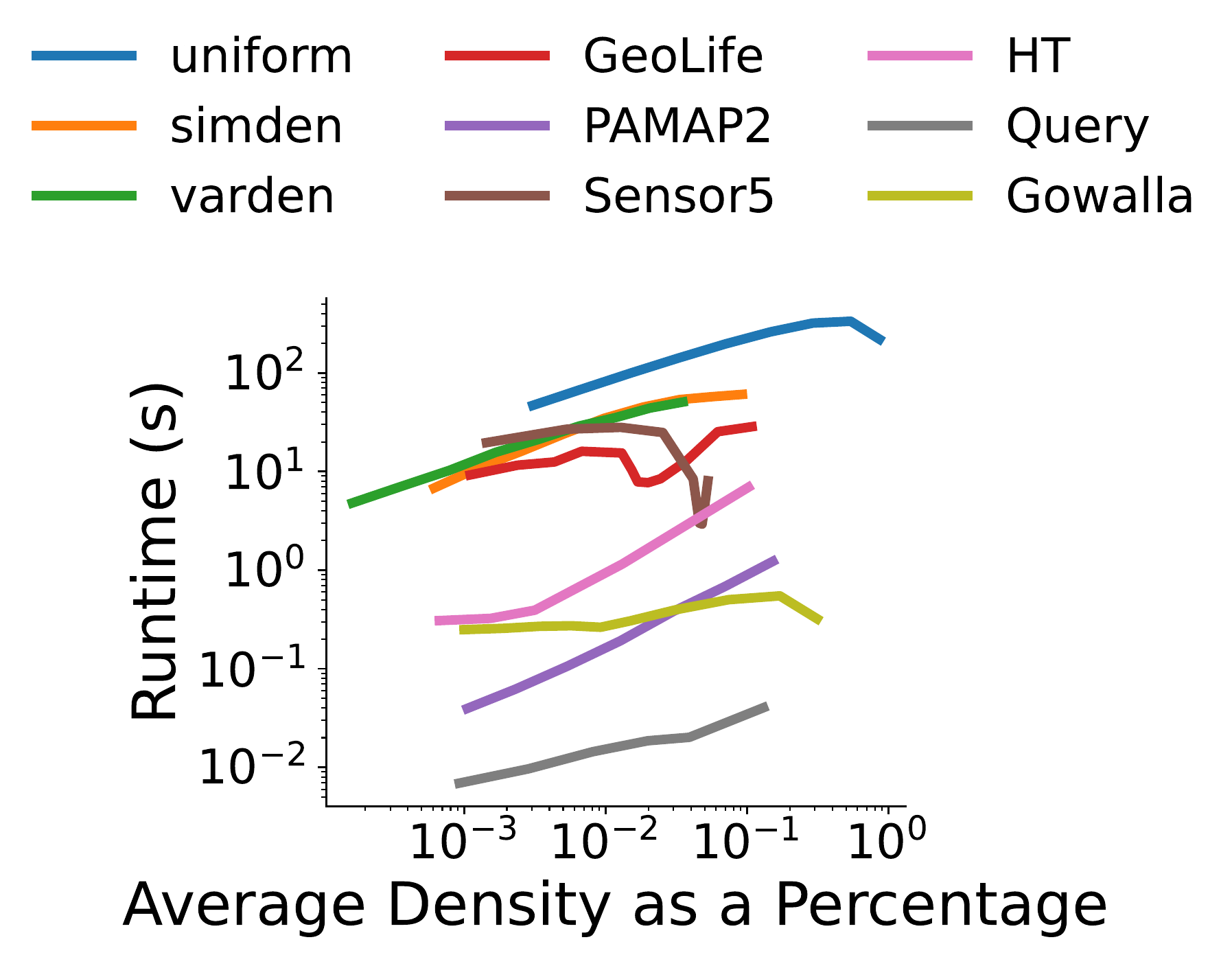}}}
    \quad
    \subfloat[Running time (seconds) of dependent point finding.\label{fig:dpc-dcut-dependent}]{{\includegraphics[width=0.64\columnwidth]{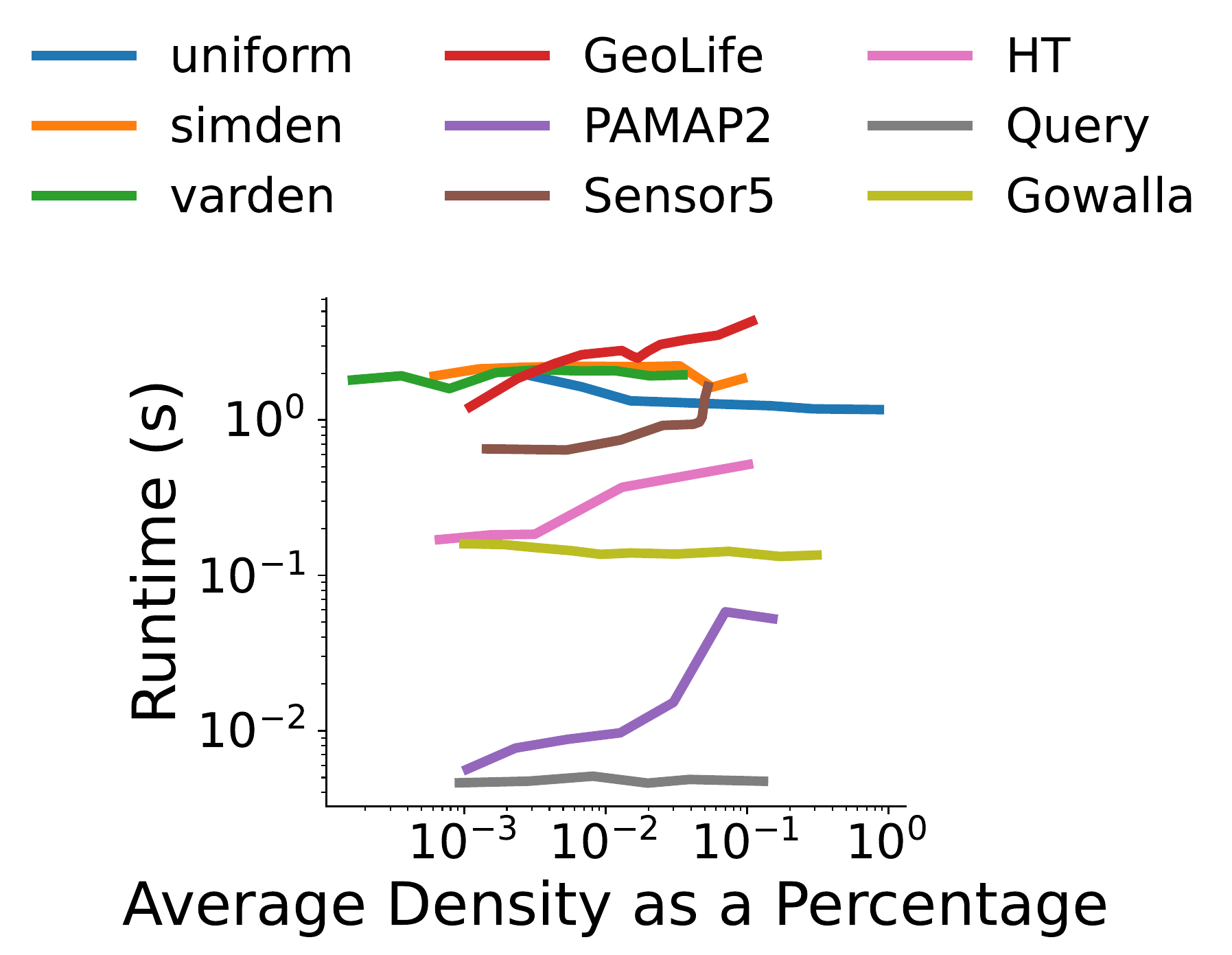}}}
    \caption{Running time on 30 cores with hyper-threading of \algname{dpc-priority}  on various data sets using different $d_\text{cut}$ values. The $x$-axis shows the average percentage of points within the $d_\text{cut}$ radius.}\label{fig:dpc-dcut} %
\end{figure*}

\section{Effect of the $d_\text{cut}$ parameter on running time}\label{sec:dcut}

\Cref{fig:dpc-dcut}  shows the running time using 30 cores with hyper-threading of our priority search \kdt based DPC algorithm on various data sets using different $d_\text{cut}$ values (other parameters are kept the same). The $x$-axis of \Cref{fig:dpc-dcut} is the average percentage of points in the data set within the $d_\text{cut}$ radius. A higher $d_\text{cut}$ value gives a higher average percentage. 

In \Cref{fig:dpc-dcut-total} and \Cref{fig:dpc-dcut-density}, we see that in general, a higher $d_\text{cut}$ value leads to a higher overall running time  and a higher running time in the density computation step. This is because a larger range query region intersects more cells of the priority search \kdt and thus leads to more work for computing the density values.

In \Cref{fig:dpc-dcut-dependent}, we see that an increase in $d_\text{cut}$ tends to also result in an increase in running time for dependent point finding, but the correlation is much weaker. The positive correlation is due to the fact that we skip dependent point computation for points with density lower than the cutoff $\rho_\text{min}$. With $\rho_\text{min}$ kept constant, a larger $d_\text{cut}$ value results in fewer noise points being skipped, and more dependent point computations being done. 

\fi

\end{document}